\documentclass[12pt,preprint]{aastex}

\begin{document}

\title{Dynamical Zodiacal Cloud Models Constrained \\ by High Resolution
Spectroscopy of the Zodiacal Light}

\author{Sergei I. Ipatov $^{a,b,*}$, Alexander S. Kutyrev $^c$, 
Greg J. Madsen $^{d,1}$, John C. Mather $^c$, \\
S. Harvey Moseley $^c$, Ronald J. Reynolds $^e$}
\affil{
$^a$ Department of Terrestrial Magnetism of Carnegie Institution of Washington,
5241 Broad Branch Road, Washington, DC, 20015-1305, USA\\
$^b$ Space Research Institute, 84/32 Profsoyuznaya st., Moscow, 117997, Russia\\
$^*$ Corresponding Author E-mail address: siipatov@hotmail.com\\
$^c$ NASA/GSFC, Greenbelt, MD 20771\\
$^d$ Anglo-Australian Observatory, P.O. Box 296, Epping, NSW 1710,
Australia\\
$^e$ Department of Astronomy, 475 North Charter st., University of
Wisconsin at Madison, Madison, WI 53706, USA}
\altaffiltext{1}{NSF Distinguished International Postdoctoral Research
Fellow}

\vspace{2cm}

Pages: 45
 
Tables: 2

Figures: 8

\newpage

{\bf Proposed Running Head:}
Dynamical zodiacal cloud models

Ipatov et al.

{\bf Editorial correspondence to:}

E-mail address: siipatov@hotmail.com

Dr. Sergei Ipatov, 22 Parkway rd., Apt. A, Greenbelt, MD 20770

\newpage

{\bf ABSTRACT}

The simulated  Doppler shifts of the solar Mg~I Fraunhofer line
produced by scattering on  the solar light by asteroidal, 
cometary, and trans-Neptunian
dust particles are compared with the shifts obtained 
by Wisconsin H-Alpha Mapper (WHAM) spectrometer.
The simulated spectra are based on the results of 
integrations of the orbital evolution of particles 
under the gravitational influence of planets, 
the Poynting-Robertson drag, radiation pressure, and solar wind drag.
Our results demonstrate that the differences in the line centroid 
position in the solar elongation
and in the line width averaged over the elongations 
for different sizes of particles are usually less than those
for different sources of dust.
The deviation of the derived spectral parameters 
for various sources of dust used in the model
reached maximum at the elongation (measured eastward from the Sun) 
between 90$^\circ$ and 120$^\circ$.
For the future zodiacal light Doppler 
shifts measurements, it is important  to pay a particular attention  to 
observing at this elongation range.
At the elongations of the fields observed by WHAM, the 
model-predicted Doppler shifts were close to each other  for several 
scattering functions considered.  
Therefore 
the main conclusions of our paper don't depend on
a scattering function  and mass distribution of particles 
if they are reasonable.
  A comparison of the dependencies of the 
Doppler shifts on solar elongation and the 
mean width of the Mg~I line
modeled for different sources of dust  with
those obtained  from the WHAM observations
shows that the fraction of cometary particles in zodiacal dust
is significant and can be dominant.
    Cometary particles originating inside Jupiter's orbit and
particles originating beyond Jupiter's orbit 
(including trans-Neptunian dust particles)
can contribute to zodiacal dust about 1/3 each,
with a possible deviation from 1/3 up to 0.1-0.2.
The fraction of asteroidal dust is estimated to be $\sim$0.3-0.5.
The mean eccentricities of zodiacal particles
located at 1-2 AU from the Sun
that better fit the WHAM observations
are between 0.2 and 0.5, with a more probable value of about 0.3.

{\bf Key Words:} Asteroids; Comets, dust; Trans-Neptunian objects; 
Spectroscopy; Zodiacal light

\newpage

\section{Introduction}

A lot of dust particles are produced by small bodies
in the solar system. The dust located within about 2 AU from the Earth 
is seen as the zodiacal light.
There are various points of view on the contributions
of asteroidal, cometary, and trans-Neptunian dust to the zodiacal 
cloud. 
The estimates of the contributions made 
in several works are summarized in Table 1.
These estimates were based on the Infrared Astronomical Satellite ({\it IRAS}) 
and {\it COBE/DIRBE} observations, on cratering rates, shape of microcraters, etc.
In the present paper, for estimates of the contributions 
we analyzed some of these observations using our 
studies of migration of dust particles produced by different small bodies.
We considered a wide range of particle masses, whereas 
some other scientists used results of calculations for
one or two sizes of particles, e.g., Liou et al. (1995) considered
9 $\mu$m diameter dust particles, and studies by Gorkavyi et al. (2000a,b)
and Ozernoy (2001) were based on 1 $\mu$m and 5$\mu$m 
particles modeling.
In our analysis for the first time 
we use the observations of velocities of zodiacal
dust particles obtained by Reynolds et al. (2004) with the
use of the Wisconsin H-Alpha Mapper (WHAM) spectrometer.

To set the stage for our work we first review some 
 published estimates of asteroid/cometa\-ry contributions to the zodiacal 
 cloud in more detail than in Table 1.    
A significant fraction of cometary dust in the near-Earth space was
proposed by Southworth (1964), Liou et al. (1995), and Zook (2001).
Based on cratering rates from an ensemble of Earth-
and Lunar-orbiting satellites, Zook (2001) estimated that the cometary
contribution to the near-Earth flux of particles is $\sim$75\%.
His conclusion was based on (1) the comparison of the
meteoroid penetration rates of the 25-$\mu$m thick cells of the
Earth-orbiting {\it Explorer 16} and {\it 23} satellites with
the penetration rate of the five {\it Lunar Orbiter} satellites
that had nearly identical cells of the same thickness and on
(2) the studies of the crater production rate on the leading edge
of the Earth-orbiting Long Duration Exposure Facility ({\it LDEF}) 
satellite, as compared to that on the trailing edge. 
For the estimates of the cometary contribution, 
Zook also used (1) Humes's (1993)
result that it takes a mean
Earth-entry velocity of about 17 km s$^{-1}$ to give
agreement with the {\it LDEF} observations and (2) 
Jackson and Zook's (1992) numerical modeling, which showed
that meteoroids originating in the main belt of asteroids
will strike the top of the Earth's atmosphere with a mean
velocity between 12 and 13 km s$^{-1}$.

$<$ Place for Table 1 $>$
\centerline{}

Grogan et al. (2001), Dermott et al. (2001), and Wyatt (2005)
suggested that at least 30\% of zodiacal
dust comes from the break-up of asteroids in order to explain formation of
dust
bands (i.e., excesses of dust at elliptic latitudes $\le$10$^\circ$
(Kelsall et al. 1998)), as dust bands alone supply as much as 30\%.
Kortenkamp and Dermott (1998)
suggested that the Earth predominately accretes asteroidal dust.
Dermott et al. (2002) concluded that most of zodiacal dust particles
can be of asteroidal origin and have eccentricities $e$$<$0.1.
Gr\"un (1994) and Gr\"un et al. (2001) considered that
zodiacal particles 
orbit the Sun at low inclination ($i$$<$30$^{\circ}$) and moderate
eccentricity ($e$$<$0.6) orbits.
Our studies presented in Section 4 
are in accordance with Gr\"un's estimates.

Nesvorn\'y et al. (2006) compared the {\it IRAS} observations
with their computer model
of the thermal emission of the Karin and Veritas family particles.
Their best-fit model results suggest that the Karin and Veritas family particles 
contribute by 5-9\% in 10-60-$\mu$m wavelengths to the zodiacal
brightness within 50$^\circ$ latitudes around the ecliptic,
and by 9-15\% within 10$^\circ$ latitudes. The high brightness 
of the zodiacal cloud at large latitudes suggests that it is mainly 
produced by particles with higher orbital inclinations than what would be
expected for asteroidal particles produced by sources in the main asteroid belt.
Based on these results, Nesvorn\'y et al. infer that asteroidal dust represents a
smaller fraction of the zodiacal cloud than previously thought
(e.g., by Dermott et al., 2001). 
They hypothesize that up to $\approx$50\% of interplanetary dust particles
measured by the {\it LDEF} may be made up of 
particles species from the Veritas and Karin families.
Based on their modelling, the disproportional contribution of Karin/Veritas
particles to the zodiacal cloud (only 5-9\%) and to the terrestrial
 accretion rate (30-50\%) suggests that the effects of gravitational
 focusing by the Earth enhance the accretion rate of Karin/Veritas particles 
relative to those in the background zodiacal cloud. Nesvorn\'y et al. (2006)
noted that the size distribution of asteroidal particles can be a
strong function of the heliocentric distance. They 
infer that the zodiacal cloud emission may be dominated by high-speed cometary 
particles, while the terrestrial impactor flux contains a major contribution 
from asteroidal sources. 
Hahn et al. (2002) concluded that, though about 80\%
of the dust particles in the sample of dust collected in the Earth's
stratosphere by U2 aircraft (Brownlee et al. 1993) have
low entry velocities consistent with asteroidal orbits,
the dust released from low-inclination Jupiter-family comets
can also  have low entry velocities, and the Earth's gravitational
focusing naturally selects for low-velocity dust over  all dust.

Sykes et al. (2004) infer that the zodiacal cloud scale height
is not a good discriminator
between asteroids and comets as the main
supply of dust because 
inclinations of some Jupiter-family comets are not very different from 
those for asteroids and
the half-width of the distribution of asteroidal  orbital inclinations
(12$^\circ$-16$^\circ$) does not differ much from the half-width
(14$^{\circ}$) at half-maximum number density for the Kelsall model.
Some models based on {\it in situ} particle detections suggest
that the inclination distribution may have a half-width as wide
as 20$^\circ$-30$^{\circ}$ (Dikarev et al. 2002).
If the latter estimate is true, then the fraction 
of zodiacal dust produced by 
long-period comets can exceed 10-20\%.
 
Landgraf et al. (2002) concluded that Comet 29P/Schwassmann-Wachmann 1 
itself is
able to provide a major fraction of the solar system dust that is
currently found between 6 and 8 AU.
Comparison of the number density of dust particles produced by
different small bodies with the observed constant density
of $\sim$10 $\mu$m particles between 3 and 18 AU (Humes 1980,
Gr\"un 1994) showed (Ozernoy 2001; Landgraf et al. 2002; Ipatov and Mather
2006) that a considerable fraction of dust
at such distances is produced by comets.
The conclusions on a considerable fraction of cometary dust
are also in agreement with
earlier studies of the dynamics of Jupiter-family comets (Ipatov and Mather
2003, 2004a-b, 2006), which showed that some former cometary objects
could get high eccentric orbits located entirely inside Jupiter's orbit
and stay in these orbits for a long time. Some of these
objects could disintegrate producing a substantial amount of dust.

Gr\"un et al. (1985) proposed that the main
contribution to the zodiacal light is from particles that range from 20 to
200 $\mu$m in diameter (for silicate particles, this range corresponds to
the ratio of the Sun's radiation pressure force 
to gravitational force $\beta$$\sim$0.002-0.02).
The cratering record on {\it LDEF}  
showed (Love and Brownlee 1993) that the cross-sectional area distribution
of particles accreted by the Earth reaches maximum at a particle 
diameter $d$$\sim$100 $\mu$m.
Smaller particles have a larger surface per unit of mass of particles.
For a fragmentation power law $n(r_p)dr_p$=$n_{\circ}r_p^{-q}dr_p$,  
the brightness of produced particles is proportional to $r_p^{2-q}$.  
Dynamical lifetimes  of particles are
usually smaller for smaller particles, 
and many particles with
diameter $d$$\sim$1 $\mu$m are relatively quickly removed from the solar system.
If we consider that a lifetime of a particle $t_{dl}$$\sim$$d$ (at 1$<$$d$$<$1000
$\mu$m, 
this assumption is in a general agreement with dynamical lifetimes of particles 
 obtained by Ipatov and Mather 2006; destruction of particles is not considered) and
$q$=3.5,
then the brightness of migrating particles with diameter $d$ is proportional to
$d$$^{-1/2}$
and the total brightness of particles with diameters between $d_1$ and $d_2$ is
proportional to $\sqrt{d_2} - \sqrt{d_1}$. These our estimates are in accordance with
that particles with $d$$<$20 $\mu$m are not main
contributors to the zodiacal light.

Dermott et al. (2001, 2002) noted that the cross-sectional area of
material in the asteroid belt should be concentrated in particles
with $d$$\sim$1000 $\mu$m and many of these large particles are
broken up by collisions before they reach a distance from the Sun $R$=1 AU.
Gr\"un et al. (1985) considered that the Poynting-Robertson drag lifetime is 
comparable to the collisional lifetime for zodiacal particles greater than 100
$\mu$m. 
Liou et al. (1996) and Moro-Martin and Malhotra (2002) studied collisional
lifetimes of interplanetary particles destroyed by interstellar dust
particles,
and the sublimation temperature of particles at several distances to the
Sun.
They found that collisional destruction is most important for
trans-Neptunian particles with $d$ between 6 and 50 $\mu$m.

Ground-based spectroscopic observations allow one to study Doppler shifts
of scattered solar Fraunhofer lines in the zodiacal light (e.g., James
1969; Hicks et al. 1974; Fried 1978; East and Reay 1984;
Clarke 1996; Clarke et al. 1996).  Analysis of
these shifts thus provides an opportunity to explore velocities of
interplanetary dust in the inner solar system.  
Reynolds et al. (2004) were the
first to obtain accurate measurements of the centroid velocities and line
profiles of the scattered solar Mg I $\lambda$5184 absorption line in the
zodiacal light, both along the ecliptic equator and at high ecliptic
longitudes.

The main goal of our paper is to compare the WHAM observations
with our models of the zodiacal dust cloud based on our
calculations of the migration of dust particles
produced by different small bodies.
The models for migration of dust particles 
and calculations of the radial velocity profile of the
scattered Mg~I line are discussed in Section 2.
Earlier we (Ipatov et al. 2005, 2006; Ipatov and Mather 2006;
Madsen et al. 2007) compared velocities corresponding to shifts of 
the Mg~I line obtained in our 
models with the WHAM observations in a short form for only a few sizes
of particles.  
Our present studies 
of the velocities  
(Section 3) 
are based on analysis of many data obtained for a wide range 
of sizes of particles and for various sources of
particles. In the papers by other authors, spectroscopic
observations of the zodiacal light have been
compared with analytical models, but this is the first comparison 
of the models based numerical integrations of migration of particles
with the observations.
We also compare in detail the mean width of the line (the end of Section 3)
and the variation of a  number density with distance from 
the Sun $R$$\le$3 AU (Section 5.1)
obtained in our model for different sources and sizes of 
particles with the results of observations, study
the typical eccentricities and inclinations of zodiacal particles 
that better fit the WHAM observations (Section 4), and discuss  
the fractions of asteroidal, cometary, and trans-Neptunian particles
in the zodiacal cloud which better satisfy various observations (Section 5).

\section{Model}  

Our studies of the Mg~I line shifts (see Section 3) use the results of following the
orbital evolution of about 15,000 asteroidal, cometary, and
trans-Neptunian dust particles under the gravitational influence of
planets, the
Poynting-Robertson drag, radiation pressure, and solar wind drag.
Results of some of these integrations were presented by Ipatov et al.
(2004) and Ipatov and Mather (2006, 2007)
(our recent papers can be found on astro-ph and on
\url{http://www.astro.umd.edu/$\sim$ipatov} or
\url{http://www.dtm.ciw.edu/ipatov}), but other problems
(mainly the probabilities of collisions of particles with the terrestrial
planets) were considered. In this section we describe
models used in our studies of the migration of dust particles
and for calculation of the scattered line profile.

\subsection{
Sources and sizes of model particles}

The initial positions and velocities of asteroidal particles (`ast' runs)
used in our
models were the same as those of the first $N$ numbered main-belt
asteroids (JDT 2452500.5), i.e., dust particles were assumed to leave the
asteroids with zero relative velocity
(in Section 2.2 
we discuss why we can make such assumption). The initial positions and
velocities of the trans-Neptunian particles ({\it tn} runs) were the same
as those of the
first $N$ trans-Neptunian objects (TNOs) (JDT 2452600.5). 
These objects had semi-major axes less than 48 AU and 
eccentricities less than 0.35.

The initial positions and
velocities of cometary particles were the same as those of Comet 2P/Encke
($a_\circ$$\approx$2.2 AU, $e_\circ$$\approx$0.85,
$i_\circ$$\approx$12$^\circ$), or
Comet 10P/Tempel 2 ($a_\circ$$\approx$3.1 AU, $e_\circ$$\approx$0.526,
$i_\circ$$\approx$12$^\circ$), or Comet 39P/Oterma ($a_\circ$$\approx$7.25
AU,
$e_\circ$$\approx$0.246, $i_\circ$$\approx$2$^\circ$), or test long-period
comets
($e_\circ$=0.995 and $q_\circ$=$a_\circ$$\cdot$(1-$e_\circ$)=0.9 AU
or $e_\circ$=0.999 and $q_\circ$=0.1 AU, $i_\circ$ varied
from 0 to 180$^\circ$ in each calculation, particles produced at perihelion;
these runs are denoted as {\it lp} runs),
or test Halley-type comets
($e_\circ$=0.975, $q_\circ$=0.5 AU, $i_\circ$ varied
from 0 to 180$^\circ$ in each calculation, particles launched at perihelion;
these runs are denoted as {\it ht} runs).
The number of prograde Halley-type comets is greater than
the number of retrograde
Halley-type comets, but in {\it ht} runs we considered a uniform 
distribution in $i_\circ$
in order to study the role of variation in $e_\circ$  in comparison
with {\it lp} runs.
We considered Encke particles launched near perihelion (runs denoted as 2P),
near aphelion (runs denoted as `2P 0.5t'),
and when the comet had orbited for $P_a/4$ after perihelion passage, where
$P_a$ is the period of the comet (runs denoted as `2P 0.25t').
Calculations for particles originating from Comets 10P/Tempel 2 and 
39P/Oterma are denoted as 10P and 39P runs, respectively.
Note that for the same initial coordinates and velocities,
initial semi-major axes and eccentricities of dust particles
depend on $\beta$ and differ
from those of parent bodies, but inclinations are the same (Burns et  al.
1979).
All orbital elements considered in the paper take this effect into account.

For cometary particles (exclusive for {\it lp} and {\it ht} runs,
in which all particles launched  
in perihelion),
the initial value of time $\tau$ after passing perihelion
was varied (Ipatov and Mather 2006) for different particles with a step $d \tau$=1 day
or $d \tau$=0.1 day near the actual value of $\tau$ for the comet
 (true anomaly can be considered instead of $\tau$). Comet
10P/Tempel 2 is an example of a typical Jupiter-family comet moving inside
Jupiter's orbit; Comet 39P/Oterma moves outside of Jupiter's orbit.
Comet 2P/Encke is the
only known high-eccentricity comet with aphelion distance $Q$$<$4.2 AU,
but there could be smaller cometary objects in such orbits. 
 Ipatov and Mather (2003, 2004a-b) obtained that some Jupiter-crossing objects 
can get orbits entirely located inside Jupiter's orbit and move in such orbits 
for millions or even hundreds of millions of years. Probably most of such 
objects disintegrate during such times and produce smaller objects.
Comet Encke comes close to the Sun and produces a lot of dust (Lisse et al. 2004).

In our calculations for asteroidal and cometary particles, the values of $\beta$,
the ratio of the Sun's radiation pressure force to gravitational force,
varied from $\le$0.0004 to 0.4.  Burns et al. (1979) obtained
$\beta$=$0.573 Q_{pr}/(\rho s)$, where $\rho$ is the particle's density in
grams per cubic centimeter, $s$ is its radius in micrometers, and $Q_{pr}$
is the radiation pressure coefficient, which is close to unity for
particles larger than 1 $\mu$m. For silicates at density 
of 2.5 g cm$^{-3}$,
the $\beta$ values equal to 0.004, 0.01, 0.05, 0.1, and 0.4 correspond to 
particle
diameters $d$ of about 120, 47, 9.4, 4.7, and 1 microns, respectively. For
water ice, $d$ is greater by a factor of 2.5 than that for silicate
particles. The orbital evolution of dust particles 
was studied by us for a wider
range of masses (including particles up to several millimeters)
than in most papers by other authors
(e.g., Dermott et al. 2001, 2002; Gor'kavyi et al. 1997, 1998; Gorkavyi et al. 
2000a-b; Grogan
et al.
2001; Kortenkamp and Dermott 1998; Liou et al. 1995, 1996, 1999; Liou and
Zook 1999; Moro-Martin and Malhotra 2002, 2003; Ozernoy 2001; Reach et
al. 1997). 
Most scientists considered particles with diameter $d$$<$50 $\mu$m.
Wide range of diameters  
was considered only by Nesvorn\'y et al. (2006) and Kehoe et al. (2007) 
for asteroidal particles from the Veritas and Karin families.

\subsection{Integration of the motion of dust particles}

In our integrations we took into account the gravitational influence of
planets (excluding Pluto for asteroidal and cometary particles), the
Poynting-Robertson drag, radiation pressure, and solar wind drag.
As Liou et al. (1999) and
Moro-Martin and Malhotra (2002), we assume the ratio of solar wind drag to
Poynting--Robertson drag to be 0.35.
According to Gr\"un et al. (2000), the Lorentz
force is comparable to solar gravitational interaction for particles of
$d$$\sim$0.1 $\mu$m at 1 AU and of $d$$\sim$1 $\mu$m at 50 AU from the Sun.
Interstellar particles dominate among such small particles, but they are not
significant contributors to the zodiacal light. Since we
considered mainly larger interplanetary particles, we did not include
the Lorentz force in our modeling.

Migration of dust particles was integrated using the Bulirsh-Stoer method
(BULSTO) with the relative error per integration step less than $10^{-8}$.
The BULSTO code in the SWIFT integration package (this package also includes 
a symplectic code) by Levison and Duncan (1994) was modified to
include the additional forces of radiation pressure, Poynting-Robertson
drag, and solar wind drag. The integration continued until all of the
particles either collided with the Sun or drove away to 2000 AU from the 
Sun.
For small $\beta$, considered time intervals exceeded 50-80 Myr (240 Myr for
trans-Neptunian particles). In each calculation (with a fixed source of 
particles
and $\beta$=const) we took $N$$\le$250 particles, because for $N$$\ge$500
the computer time per calculation for one particle was several times
greater than for $N$=250. The total number of particles in several tens of
runs 
was about
15,000. In our calculations, orbital elements were stored with a step  
$d_t$ of
20 yr for asteroidal and cometary particles and 100 yr for trans-Neptunian
particles during all considered time intervals.  The stored orbital
elements of all particles during their dynamical lifetimes were then used
in our studies 
presented in the next sections.

The largest asteroids and TNOs do not represent accurate orbital distribution
of bodies producing dust in the asteroid and trans-Neptunian belts, 
but for our conclusions we do not need to consider
more accurate distributions than those we used. 
For example, for solar elongation  
60$^\circ$$\le$$\epsilon$$\le$180$^\circ$ 
at average initial eccentricity $e_\circ$ of particles originating inside 
Jupiter's orbit equal to 0.15 and 0.5 ({\it ast} and 10P runs), 
in Section 3 we obtained the shift of spectra received at the Earth (from the 
solar spectrum) in the same direction (to blue). 
At $e_\circ$$\ge$0.85 the shift was in another direction. 
Therefore for conclusions of the present 
paper, the difference in $e_\circ$ between 0.15 and 0.2 or between 0.4 and 0.5 is 
not essential. Considered parent comets show examples of comets moving inside 
Jupiter's orbit at two different eccentricities (0.53 and 0.85), a comet outside of 
Jupiter's orbit, and comets moving with $e_\circ$$>$0.97. 
In the present paper we find 
general dependence of spectral shift  on eccentricity.
Thus for our  
estimates we do not need to make
integrations for many different comets.

Each integration was made for a fixed size of particles. We did not study
mass distribution of particles,  
but as it is discussed in Section 5.2, 
for the conclusions made in the present paper, we do not need to know accurate
mass distributions of particles. 
Therefore we did not consider 
destruction of colliding particles. The destruction affects mainly
lifetimes of particles and their size distributions at different distances
from the Sun. It can change the distribution of the particles' orbital 
elements during their migration via the zodiacal cloud, but,
in our opinion, these changes cannot affect the conclusions of the present paper, 
because these conclusions will not be changed even if real 
mean eccentricities in a run will differ by a factor  of up to 1.5
from the values obtained for our model without destruction. 
Future models, which will consider the size distributions and 
destruction of particles, will allow one to make more accurate estimates of 
the fractions of zodiacal particles of different origin
than those presented in the present paper.

Planets were assumed to be material points.  However, using orbital
elements obtained with a step $d_t$, Ipatov and Mather (2006, 2007) 
calculated
the mean probability of a collision of a particle with the terrestrial
planets during the particle dynamical lifetime.
Later we considered the probabilities of collisions of migrating particles
with the giant planets.
For most calculations, the total probability $p_{all}$ of collisions of a
particle with all planets
during a dynamical lifetime of the particle was small (less than 0.01).
Only for {\it tn} and 39P particles, $p_{all}$ could exceed 0.01.
For trans-Neptunian particles,
the probabilities $p_{jn}$ of collisions of particles with all giant
planets during dynamical lifetimes of the particles
were about 0.15 at  $\beta$$\sim$0.002-0.01
and did not exceed 0.05 at $\beta$$\ge$0.05. At $\beta$ equal to
0.002, 0.01 and 0.05  for {\it tn} runs, the main contribution to $p_{jn}$
(0.9$p_{jn}$, 0.5$p_{jn}$, and 0.5$p_{jn}$) was due to Neptune,
Jupiter, and Saturn, respectively.
For Comet 39P particles, the values of $p_{all}$$\approx$$p_{jn}$ were
about 0.018, 0.044, and 0.017
at $\beta$ equal to 0.0001, 0.001, 0.01, respectively, and
Jupiter's contribution was about 85\%.
For particles produced by asteroids and other
comets, the values of $p_{all}$  were smaller than
those for trans-Neptunian and Comet 39P particles,
and the main contribution to $p_{all}$ was due to Jupiter
if particles reached Jupiter's orbit (for some runs for all particles,
aphelion distance $Q$$<$5 AU).
For example, for {\it ht} particles we obtained 
$p_{all}$$\approx$$p_{jn}$$\le$0.004.

As $p_{all}$ is relatively small, then
even if some particles actually collide with planets,
the distribution of particles over their
orbital elements during their dynamical lifetimes will be practically the
same as in the model for which planets are considered as material points,
and the average dynamical lifetime of particles
in our calculations usually will be greater than
the actual value $T_{av}$ by less than $p_{all}$$T_{av}$.
  The probability of a collision of a particle with Jupiter is smaller than 0.1 
for trans-Neptunian particles and can be much smaller for other particles. 
For a small number of particles ($N$$\le$250), our approach
can give better estimates of the probabilities 
of collisions of particles with planets 
than direct integration of the collisions, 
especially in the cases when the expected
number of collisions of all $N$ particles does not exceed 1.
For initial data considered, most of $N$ 
particles did not collide with planets, 
and even if some particles in our calculations
changed their eccentricities at too close encounters with planets -- material
points, the main contribution to variations in mean orbital elements was from
particles that had not such very close encounters with planets
that actually could result in collisions. 
Accurate values of mean eccentricities and inclinations are not
needed for conclusions made in the present paper.
Therefore we expect that our considered model does not change the distribution of
orbital elements of particles that enter the zodiacal cloud 
in such a way that it can influence the conclusions of the paper.

In our calculations we considered particles leaving the parent bodies
with a zero velocity. Actually such velocities have nonzero values,
but it does not affect the conclusions of the present paper
because relative velocities of particles produced by  
asteroids, TNOs, and comets are small compared to their orbital
velocities and even to differences between orbital and circular velocities. 
Results of studies of particles ejected from Comet Tempel 1
showed (e.g., Jorda et al. 2007, Ipatov and A'Hearn 2006) that even 
for the collision of the Deep Impact (DI)
spacecraft with the comet at a velocity of 10 km s$^{-1}$, 
relative velocities of most ejected particles did not
exceed 200-300 m s$^{-1}$.
Typical collisional velocities in the main asteroid belt are 
about 5 km s$^{-1}$ (Bottke et al. 1994) and are 
smaller than the  velocity of the DI collision,  
so typical
relative velocities of dust particles originating from asteroids
will be smaller than 200 m s$^{-1}$. 
Gombosi et al. (1985) and Sekanina (1987) concluded that
the initial velocities of particles relative to a comet are
less than 1 km s$^{-1}$. 
Each our calculation was made for
various parent asteroids (or TNOs), so if we will
consider a nonzero distribution of relative velocities of dust
particles, the final distribution of orbital elements of produced
particles will be practically the same as that for a zero relative
velocity.

We studied the model for which a particle
collides with the Sun when perihelion distance 
of its orbit reached the radius of the Sun.
For most considered runs, exclusive of some {\it lp}, {\it ht}, and
2P runs, for the above model, dynamical lifetimes of particles are practically the
same as those for the model in which we consider direct collisions of
particles with the Sun (for the latter model, a step of integration could be 
greater than radius of the Sun).

\subsection{The Scattered line profile}

We calculated how the solar spectrum was changed after the light had been 
scattered by the dust particles and observed at the Earth. This was carried
out by first considering all orbital elements of dust particles during a
single run, which were stored in computer memory with a step $d_t$.
Based on these stored orbital elements, we
calculated velocities and positions of particles and the Earth during the
dynamical lifetimes of the particles. For each pair of positions of a
particle and the Earth, we then calculated many ($\sim$10$^2$-10$^4$,
smaller values are for larger maximum dynamical lifetimes of 
particles) different positions of the particle and the Earth
during the period $P_{rev}$ of revolution of the particle around the Sun,
assuming that orbital elements do not vary during $P_{rev}$. 
The model, which is based on all positions and velocities of dust
particles during their dynamical lifetimes, represents the zodiacal dust cloud
for the case when small bodies continuously produce dust at a constant
rate along their orbits. We did not consider seasonal effects and jumps 
in production of dust. The model considered allows one to study the
main differences between spectra corresponding to particles produced
by asteroids, comets, and trans-Neptunian objects.

The choice of a scattering function was based on analysis of dependences
of scattering functions on angles $\theta$ and $\epsilon$ (see below)
and wavelength presented
in several papers (e.g., Giese 1963; Giese and Dziembowski 1969; Leinert
1975; Leinbert et al. 1976; Weiss-Wrana 1983; Hong 1985; Lamy and Perrin
1986), which mainly followed the Mie theory for scattering. The scattering
function depends on the composition of particles, their sizes, and other
factors.  
However, we considered three simple scattering functions: (1) 
$g$=$g_{\theta}$=$1/\theta$ for $\theta$$<$$c_{\theta}$, and 
$g_{\theta}$=$1+(\theta -c_{\theta})^2$
for $\theta$$\ge$$c_{\theta}$, where $\theta$ is the angle between the Earth and the
Sun, as viewed from the particle, in radians, and $c_{\theta}$=$2\pi$/3 radian; (2)
besides the above dependence of $g_{\theta}$ on $\theta$, the same dependence 
$g_{\epsilon}$ on elongation
$\epsilon$ was considered ($g$=$g_{\theta}$$\cdot$$g_{\epsilon}$), where $\epsilon$ is 
the angle between the particle and the Sun, as viewed from
the Earth (eastward from the Sun); (3) isotropic scattering ($g$=1).  
For all
three functions, the intensity $I$ of light that reaches the Earth was
considered to be proportional to $\lambda^2$$\cdot$$(R \cdot r)^{-2}$, where 
$r$ is the distance between the particle and the Earth, $R$ is the distance
between the particle and the Sun, and $\lambda$ is the wavelength of
light. Since we considered the scattering near a single spectral line,
wavelength $\lambda$ in our calculations was essentially a constant. Except 
for lines of
sight close to the Sun, these three scattering functions give virtually
the same results (see Section 3).  Since the differences between the
scattering
functions that we considered were much greater than the differences
between scattering functions presented in the publications cited above, we
conclude that we need not worry about the precise form of the scattering.

For each particle position, we calculated $r$, $R$, 
and projections $v_{ps}$ and $v_{ep}$ of velocities of 
particles relative to the Sun and the Earth
on the lines of sight from particles to
the Sun and the Earth, respectively
(a projection of velocity is positive if a corresponding
distance increases).
These parameters and the scattering
function $g$ were then used 
to calculate zodiacal light spectrum 
(using brightness integral) as observed from the Earth. 
The line of sight is characterized by 
$\epsilon$ and by its inclination $i$ above the ecliptic plane.
Particles along the line of sight within
the beam of diameter 2$^\circ$ (Fig. 1) or 2.5$^\circ$ 
(other figures) were considered.
In each calculation, all particles had the same size
(i.e., the same $\beta$), the same scattering properties,
and the same source (e.g., asteroidal).
The main steps of the calculation of a model spectrum
are the following.
For all positions and relative velocities of particles,  
at different values of $\lambda$, we calculated the intensity of light 
that reaches the Earth after solar light has been scattered 
by a particle at considered positions and velocities of the
particle and the Earth.
For these calculations, we considered the Doppler shift of 
$\lambda$ ($d_\lambda$=$\lambda$($v_{ps}$+$v_{ep}$)/$v_l$,
where $v_l$ is the speed of light)
and the known intensity of light vs. $\lambda$ for the solar spectrum and 
supposed that the intensity of the light received at the Earth
 is proportional to $g \cdot \lambda^2/(r \cdot R)^2 $.  
We calculated the brightness integral using the particle 
distribution along the beam, provided by our model. 
The observations were done at one specific epoch, but the model accumulates 
the scattered light from particles at many different epochs.
In our opinion, variations of actual spectrum with epoch usually 
do not exceed the differences in spectra obtained in our runs 
at different masses of particles and may be within the accuracy of observations.
Spectra for different sources of dust considered in the present
paper differed significantly (see Section 3). 
Therefore, if we had observations at different epochs,
it would not affect conclusions of the present paper.

$<$ Place for Figure 1 $>$
\centerline{}

\section{Variations in Solar Spectrum Caused by Scattering by Dust
Particles}

Figure~1 shows sample spectra of scattered Mg~I $\lambda$5184 line obtained
from our calculations toward sightlines in the antisolar
direction (Fig.~1a)   
and toward the ecliptic pole (Fig.~1b).   
The spectra 
consist of intensity vs. wavelength shift $\Delta \lambda$ with respect to
5183.62 Angstrom. The thinnest
line in Fig. 1 denotes the initial (unscattered) solar spectrum. 
The plots in the figure are presented for the scattering function (2), but
the lines are practically the same for three different scattering functions
considered.
In the figure legend, the number 0.2 or 0.05  
denotes $\beta$, and `180' in Fig. 1a
denotes solar elongation $\epsilon$ (in degrees).
The WHAM observations are presented by crosses. 
These observations 
and all other plots in Fig.~1 
were stretched vertically so that the minimum in the
line was at approximately the same depth as that for the initial solar
spectrum. The continuum levels were also normalized to 1. 
Similar plots at $\epsilon$=90$^\circ$ and $\epsilon$=270$^\circ$ for zero
inclination above the ecliptic plane were presented by Ipatov et al. (2005).
Unlike results by Clarke et al. (1996) and Clarke (1996)
who considered spectrum near 4861 Angstrom ($H_{\beta}$ line of hydrogen),
our modeled spectra don't exhibit strong asymmetry.
Ipatov et al. (2005; astro-ph/0608141) 
similarly found that minima in the plots of dependencies
of the intensity of light on its wavelength near 5184 Angstrom are not as
deep as those for the initial solar spectrum.
Note that 
in the paper by Clarke et al. (1996) elongation 
is measured in the opposite clockwise direction than in our present paper,
so our 90$^\circ$ corresponds to 270$^\circ$ in their paper.

At the North Ecliptic Pole, the calculated spectrum was
shifted slightly to the left (to the blue) 
relative to the solar spectrum for asteroidal
particles and slightly to the right (to the red) for particles originating
from Comet 2P/Encke (Fig. 1b).
These shifts may be due to small asymmetries in
the model particle distributions with respect to the ecliptic plane.
The spectra of Comet 10P and Comet 39P particles and those
from long-period comets were
very similar to each other.  For cometary particles, the line profile has
a flatter bottom than that for asteroidal particles, but it was not as 
wide as the observed spectrum presented by Reynolds et al. (2004).
None of our model runs matched the large width of the observation toward
the ecliptic pole.  This issue will not be addressed in this paper, but
will be a topic for future investigation and may need additional,
more accurate observations.

Using the model spectrum similar to that presented in Fig. 1a, 
we determined the 
shift $D_{\lambda}$ of the model spectrum  with respect to the 
solar spectrum by comparing line centroids, for a number of lines 
of sight at different solar elongations $\epsilon$. 
Based on $D_{\lambda}$, we calculated `characteristic' velocity 
$v_c$=$v_l \cdot D_{\lambda}/\lambda$, where
$v_l$ is the speed of light and $\lambda$ is the mean wave length of 
the line. The plot of $v_c$ 
vs. the solar elongation $\epsilon$ along the ecliptic plane is
called the `velocity-elongation' plot.
      The Doppler shift of the line centroid in the zodiacal spectrum 
with respect to unshifted solar line depends on many properties of 
the zodiacal cloud, such as dust spatial distribution, particle sizes, 
velocities and their dispersion and scattering function. The resulting 
zodiacal light spectrum is defined by what is usually referred to in 
the literature as brightness integral. Inverting brightness integral 
and solve for the real dust particles velocities along the line of 
sight is not a trivial task (e.g. Schuerman 1979) and can be quite 
challenging. Note that $v_c$ corresponding to the Doppler shift of 
the line centroid is not a velocity that can be attributed directly 
to some particular group of particles, but merely a compound 
parameter for the model verification and its comparison with the 
observational data.
The value of $D_{\lambda}$ depends on values of $r$, $R$, $v_{ps}$, $v_{ep}$, 
and $g$ for many dust particles. This dependence is caused by that for 
construction of each plot similar to that in Fig. 1a we need to consider all
particles in the beam at a given solar elongation $\epsilon$, and
for each particle we need to calculate the intensity and the 
Doppler shift in zodiacal light spectrum observed from the Earth.
The intensity depends on $r$, $R$, and $g$, and the shift depends
on velocities $v_{ps}$ and $v_{ep}$ (see Section 2.3). For 
calculation of one value of $D_{\lambda}$, we need to know all values
of intensity vs. $\lambda$ (near $\lambda$=5183.62 Angstrom)
in a plot similar to Fig. 1a, to find the center of mass of the 
area located above the absorption line curve and under the projected level 
of continuum (in Fig. 1 the level is equal to 1),
 and to calculate the difference between the $\lambda$ coordinate
of this center of mass and that for the corresponding center 
of mass for the solar light.
    Comparison with the results of WHAM data was done in 
velocity shift $v_c$ rather than in wavelength change $D_{\lambda}$
making it consistent with generally accepted in studies 
of the zodiacal light Doppler shift.

 `Velocity-elongation' plots are presented in Figs. 2-4. 
For plots 
marked by $c$, we considered the shift of the centroid (the `center of
mass' of the line), while `velocity-elongation' curves marked by $m$ denote
the shift of the minimum of the line. `Velocity-elongation' plots for
different scattering functions are denoted as $c1$ and $m1$ for the
scattering function 1, as $c2$ and $m2$ for the function 2, and as $c3$
and $m3$ for the function 3. The lines in Fig. 1 are nearly symmetric,
so the results for `c' and `m' in Fig. 2 differ only a little.  
In Fig. 2 comparison of plots is presented for two runs, but
similar comparison was made 
and similar results were obtained 
for other sources of particles (see e.g. astro-ph/0608141). 
In Figs. 3-4 
the results were obtained using only the second scattering function.
  The values of scattering function are large at elongation close to 0. 
Therefore the absolute values of velocities in Figs. 2-4   
are large for 
small $\epsilon$. Velocity changes a sign  
at $\epsilon$=0.
 
$<$ Place for Figures 2-4
 $>$
\centerline{}

`Velocity-elongation' curves 
characterize all observations along the ecliptic plane
and allow one to make more reliable conclusions than the plots 
similar to Fig. 1.
Note that plots in Fig. 1a are presented for $\epsilon$=180$^\circ$, 
and for other elongations the difference between the model and 
solar spectra can be much greater than at $\epsilon$=180$^\circ$
(see Figs. 3-4).   
The details of the model spectra depend on $\epsilon$, $\beta$, $i_\circ$, 
and
the source of particles.
The `velocity-elongation'
curves obtained for different scattering functions were similar at
30$^\circ$$<$$\epsilon$$<$330$^\circ$ (Fig. 2); though the difference was
greater for directions close to the Sun. Some  
not smooth parts of the curves are
caused by small number statistics for these calculations. 
A comparison of the observed `velocity-elongation' curve with those
obtained from our model for dust particles of different sizes (i.e.,
at different values of $\beta$) produced by asteroids, comets (2P/Encke,
10P/Tempel 2, 39P/Oterma, long-period, Halley-type), and trans-Neptunian 
objects
allowed us to draw some conclusions about the sources of zodiacal dust
particles. Asteroidal, trans-Neptunian,  Comet 2P, and {\it lp} particles
populations
produce clearly distinct model spectra of the zodiacal light.
The curves for {\it lp} and {\it ht} runs did not differ much from each other.
The main contribution to the zodiacal light is from particles
at $\beta$$\sim$0.002-0.02 (Gr\"un et al. 1985). 
Therefore particular attention must be paid to the curves
at such values of $\beta$.
   At 90$^\circ$$\le$$\epsilon$$\le$270$^\circ$,
the projection of a velocity of a prograde particle
on the direction of Earth's velocity
is usually less than the velocity of the Earth
(but has the same sign) because the particles
are located farther from the Sun than the Earth. The difference
between the projection of a velocity of a particle 
on the direction of the Earth's velocity and
the Earth's velocity has different
signs at $\epsilon$$=$90$^\circ$ and $\epsilon$$=$270$^\circ$.
Therefore in most cases, velocities corresponding to shifts in the
Mg~I line are  
positive at $\epsilon$=90$^\circ$ and negative at
$\epsilon$=270$^\circ$.
The differences between `velocity-elongation' curves for several sources of
dust reached its maximum
at $\epsilon$ between 90$^\circ$ and 120$^\circ$ (Figs. 3--4).  
For future
observations of velocity shifts in the zodiacal spectrum, it will be
important to pay particular attention to these elongations.

$<$ Place for Table 2 $>$
\centerline{}

We consider the amplitude of `velocity-elongation' curves as
$v_a$=($v_{\max}$-$v_{\min}$)/2, where $v_{\min}$ and $v_{\max}$
are the minimum and maximum values of velocities at
90$^\circ$$\le$$\epsilon$$\le$270$^\circ$.
The observational value of $v_a$ is about 12 km s$^{-1}$ 
(if we smooth the curve).
For several dust sources, the characteristic values
of $v_a$, $v_{\min}$, and $v_{\max}$ are presented in Table 2.
Mean eccentricities $e_z$ and mean inclinations $i_z$ at distance
from the Sun 1$\le$$R$$\le$3 AU are also included in the table.
These mean values were calculated on the basis of
the orbital elements of migrating particles stored with a time 
step $d_t$$\sim$20-100 yr. 
Our calculations showed that the main contribution to the brightness
of a dust cloud observed at the Earth is from particles located at $R$$<$3 AU, and
for most of the runs  more than a half
of the brightness is due to the particles located at a distance from the Earth
$r$$\le$1 AU.  
Thus, since only
positions of particles at $R$$\ge$1 AU are used for calculation of the
brightness of particles at elongation 90$^\circ$$\le$$\epsilon$$\le$$270^\circ$, 
if it is not
mentioned specially, the mean eccentricities $e_z$ and orbital inclinations
$i_z$ refer to 1$\le$$R$$\le$3 AU.
 Particular attention was paid to eccentricities and inclinations
at 1$\le$$R$$\le$2 AU.

For asteroidal dust, the `velocity-elongation' curves had
lower amplitudes than the observations (Fig. 3a-b).  The plots obtained
at different $\beta$ differed little from each other, especially at
$\beta$$<$0.01. For Comet 10P particles, the `velocity-elongation' 
amplitudes
were also lower than that for the observations (Fig. 3c-d). The difference
between the curves obtained for Comet 10P particles at different $\beta$ was
greater than that for asteroidal particles, but was usually less than that
for other sources of particles considered. On the other hand, the
`velocity-elongation' curve corresponding to particles produced by Comet
2P/Encke have slightly larger amplitudes than the observational curve (Fig.
4a-b).
The velocity amplitudes $v_a$ for
particles originating from long-period and Halley-type comets
are much greater than those for the observational curve (Fig. 4f).  
Therefore perhaps a combination of Comet 2P (and/or {\it lp} and {\it ht})
dust particles and asteroidal (and/or Comet 10P)
particles could provide a result that is close to the observational
`velocity-elongation' curve.

The orbit of Comet 39P/Oterma is located outside of Jupiter's orbit, 
but inside Saturn's orbit. Studies of the
migration of Comet 39P particles thus give some information about the
migration of particles originating beyond Neptune's orbit that have
reached 7 AU from the Sun. 
For Comet 39P particles and
0.01$\le$$\beta$$\le$0.2 at 60$^\circ$$<$$\epsilon$$<$$150^\circ$,
`velocity-elongation' amplitudes were smaller than the
observed amplitudes (Fig. 4c),  
while for $\beta$$\le$0.004, the curves more closely match the
observations (Fig. 4d). 
For such small $\beta$, only a small number of
particles entered inside Jupiter's orbit, and statistics were poor. The
distribution of particles over their orbital elements could be somewhat
different if we had considered a greater number $N$ of particles in one
run, but the difference between the curves obtained at different $N$ will
not be more than the difference between the curves obtained at adjacent
values of $\beta$ presented in the figure. At $\beta$=0.0001,
Comet 39P particles
spent on average more time in the zodiacal cloud than at other $\beta$, 
but it was due mainly only to one particle, which during about 9 Myr moved
inside Jupiter's orbit before its collision with the Sun.  In reality 
such particle could sublimate or be destroyed in collisions during its
dynamical lifetime. Small number statistics may also be responsible for
the spread of results for {\it tn} runs 
in Fig. 4e, as only a few trans-Neptunian particles
in each run (at fixed $\beta$) entered inside Jupiter's orbit. If there
had been a greater number of particles in the {\it tn} runs, the differences
between the corresponding plots would probably have been smaller.

'Velocity--elongation' curves for asteroidal and Comet 10P  
particles are located below the observational curve,  
and those for Comet 2P particles are located above
the observational curve (Figs. 3-4). 
Therefore a combination of
different sources of particles could give a zero vertical shift.
For trans-Neptunian 
particles, velocity amplitudes $v_a$ were greater than
the observational values at $\beta$$\ge$0.05 and were about the same at
$\beta$=0.01 (Fig. 4e).  
Therefore together with high-eccentricity cometary particles, trans-Neptunian
particles can compensate small values of $v_a$ for asteroidal and Comet 10P
particles.
The observational curve was mainly
inside the region covered by trans-Neptunian curves obtained for different
$\beta$ (and by Comet 39P curves at $\beta$$\le$0.004), but at
180$^\circ$$<$$\epsilon$$<$270$^\circ$ it was mainly above
the trans-Neptunian curves.

Observations by Reynolds et al. (2004) also provided the FWHM (full width
at half maximum, i.e., the {\it x}-width at $y$=($y_{\min}$+$y_{\max}$)/2)
of the Mg~I line in the zodiacal light. 
 As Reynolds et al.,  we consider FWHM in km s$^{-1}$.
The relation between the width $\Delta \lambda_w$ of the spectrum line 
and a corresponding velocity $v_w$ is the following:
$\Delta \lambda_w / \lambda = v_w/v_l$, where  $v_l$ is the speed of light
and $\lambda$ is the mean wave length of the line.
The values of $v_w$ obtained at observations varied from 65.6 to 
87.2 km s$^{-1}$, and most of them were between 70 and 80 km s$^{-1}$, 
with a mean value of 76.6 km s$^{-1}$. For close values of $\epsilon$,
observational values of $v_w$ can differ considerably, e.g.,
$v_w$ equaled 65.6 and 83.9 km s$^{-1}$ at $\epsilon$ equaled 
to 179.3$^\circ$ and 174.2$^\circ$, respectively.
Most of the values of FWHM obtained in our models (exclusive for 2P runs)  
are inside the range of observational values (see astro-ph/0608141).
At $\beta$$\le$$0.004$ for particles originating from Comet 2P/Encke, the 
width is greater at $\epsilon$$\approx$220$^\circ$ than at
$\epsilon$$\approx$45$^\circ$.  
Such dependence of FWHM on $\epsilon$ is not consistent with the observations,
and therefore the
contribution of particles similar to Comet 2P particles to the zodiacal light 
cannot be considerable.
For other sources of dust at $\beta$$\le$$0.02$, the
width is approximately independent of $\epsilon$ at
30$^\circ$$<$$\epsilon$$<$$330^\circ$.  
For all considered sources of dust and diameters of particles, 
the widths become relatively
large at $-15^\circ$$<$$\epsilon$$<$15$^\circ$, where there are no
observations.

$<$ Place for Figure 5   
$>$
\centerline{}

As the observational values of FWHM differed much at close $\epsilon$
and may be due to observational uncertainty,
it is better to compare our models with the mean value of FWHM
obtained by observations.
For particles originated from asteroids, Comets 2P/Encke,
10P/Tem\-pel 2, and 39P/Oterma, and test long-period comets, the mean (at
30$^\circ$$<$$\epsilon$$<$330$^\circ$) values of FWHM are 
mainly about 74-76, 81-88,
76-77, 76-77, 73-86 km s$^{-1}$, respectively 
(the range is for various $\beta$). 
For Comet 10P and Comet 39P particles, the mean width is slightly
greater for smaller $\beta$
at 0.0001$\le$$\beta$$\le$0.1 (Fig. 5).  
For Comet 2P particles, the mean values of FWHM depend on
$\beta$ and the place of origin from the orbit (i.e.,
the initial value of true anomaly). 
For 2P particles originated at perihelion, mean values of FWHM were
greater than observational 76.6  km s$^{-1}$.
The mean values of FWHM
obtained for particles originated from test long-period comets can differ
considerably in different runs with different $\beta$.
For {\it lp} particles,
the mean width usually is even less than that for Comet 2P particles at the 
same $\beta$.
At $\beta$$\le$0.01 for {\it ht} particles, the
mean width was large ($\approx$90 km s$^{-1}$).

As it is summarized in Fig. 5,  
the mean value of FWHM for asteroidal dust 
at $\beta$$\ge$0.0004 is 
less than the 77~km s$^{-1}$ FWHM of the observations. To fit the
observations of the mean width for a combination of 
asteroidal and cometary dust, we need to
consider a greater ($>$50\%) fraction of Comet 10P and Comet 39P particles 
or a smaller ($<$50\%) fraction of
Comet 2P or {\it ht} particles in the overall dust, as points
for Comet 2P and {\it ht} particles
at $\beta$$\le$0.02 are located in Fig. 5  
farther from the observational value than those for asteroidal particles.

\section{Eccentricities and Inclinations of Zodiacal Dust Particles
that Fit 
the Doppler Shift
of Mg~I Line}

In order to understand the variations in the model line profiles with the
source and size of particles, we examined the values of mean
eccentricities and mean orbital inclinations of zodiacal dust particles.  
Analysis of the correlation between the values of mean eccentricities
and inclinations and the values of the velocity amplitudes
of `velocity-elongation' plots 
showed that, in general, these amplitudes
are greater for greater mean
eccentricities and inclinations, but they depend also on distributions of
particles over their orbital elements.

$<$ Place for Figures 6-7  
$>$
\centerline{}

The values of $e_z$, $i_z$, $v_a$, $v_{\min}$, and $v_{\max}$
(see designations in Section 3) 
for particles from different sources
are presented in Table 2 at several values of $\beta$. 
Analysis of this table and
Figs. 3-4, 6-7  
shows that at $e_z$$<$0.5 for particles originated inside Jupiter's
orbit (e.g., for particles produced by asteroids and Comet 10P),
 the velocity amplitudes $v_a$ are
usually smaller than the observed amplitude (12 km s$^{-1}$), 
while for most of runs
at $e_z$$>$0.5 (e.g., for 2P and {\it lp} runs), $v_a$ is greater than
the observed amplitude. For these data,
the WHAM observations correspond to a mean
orbital eccentricity $e_z$ of about 0.5.  However,  
the velocity amplitudes of the line depend not only on $e_z$, but also on
the distribution of all orbital elements of dust particles.
For particles migrated from outside of Jupiter's orbit ({\it tn} and 39P
runs), the mean eccentricities that satisfy the WHAM observations
can be $\sim$0.1-0.4. For example, at $\beta$=0.004 for 39P run, the
data in Figs. 4d and 5   
were not far from the observations, but mean
eccentricities at 1$\le$$R$$\le$2 AU were about 0.35.
On the other side, 39P runs with $e_z$$\sim$0.6-0.7 also fit the WHAM
observations, but such large particles ($\beta$$\le$0.001) may not be
dominant in the brightness.
For {\it tn} runs at $\beta$$\ge$0.05,
mean eccentricies at 1$\le$$R$$\le$2 AU were even between 0.1 and 0.3
(between 0.2 and 0.4 at 2$\le$$R$$\le$3 AU),
but, as it is discussed in Section 5, such particles do not
dominate in the zodiacal light.
For {\it tn} runs at $\beta$$\le$0.01, only a few particles entered inside
Jupiter's orbit, and therefore it is difficult to make any reliable
conclusion.
To summarize, we can conclude that
for an abstract model of identical zodiacal particles 
from the same source, the mean eccentricities
of zodiacal particles are between 0.3 and 0.5, and they are closer
to 0.5 if most of the particles originated inside Jupiter's orbit.

Actually particles of different sources and sizes contribute to
the zodiacal light. For example (see Section 5 for discussion
of fractions of particles of different origin),
for the model of the zodiacal cloud
consisted of 40\% of {\it ast} particles with $e_z$$\approx$0.1,
40\% of Comet 10P, Comet 39P, and {\it tn} particles with 
$e_z$$\approx$0.35,
and 20\% of Comet 2P, {\it lp}, and {\it ht} particles with 
$e_z$$\approx$0.7 (for particles originated from high-eccentricity comets,
$e_z$ is smaller than initial eccentricities),
the mean eccentricity at $R$$\sim$1-2 AU will be about 0.3.
More massive particles, exclusive for those with diameter of not more
than a few microns, usually have more eccentric orbits.

In most calculations,  
mean eccentricities of particles decrease
with $R$$<$1 AU becoming close to 0 near the Sun,
and at $R$$<$1 AU the difference in mean
eccentricities for particles of different origin is
smaller than at $R$$>$1 AU.
Therefore `velocity-elongation' curves 
corresponding to different runs presented in Figs. 3-4  
become more close to each other when $\epsilon$ becomes
more close to the direction to the Sun.

The velocity amplitudes $v_a$ also depend on inclinations because 
particles in high inclination orbits
have smaller projections of their orbital velocities on
the lines of sight from the Earth and the Sun than 
particles in orbits located near the ecliptic plane. 
 The differences between these projections and the Earth's velocity are 
greater than those for small inclinations, and a vertical component 
of a relative velocity is greater for greater $i$. 
Mean inclinations of particles in the calculations that fit the WHAM observations
do not exceed 25$^{\circ}$. For 39P and {\it tn} runs, they mainly exceed
10$^{\circ}$.
For other runs usually $i_z$$>$5$^{\circ}$.
For {\it lp} and {\it ht} particles, the values of $v_a$ are
significantly greater than those for dust particles from other sources,
and the values of $i_z$ are much greater than for other
runs.
Note that mean initial inclinations for {\it lp} and {\it ht}
runs are about 90$^\circ$ (initial orbital inclinations are distributed
uniformly between 0 and 180$^\circ$), but the mean inclinations of
migrating particles in some runs are mainly greater
(and in other runs are mainly smaller) than 90$^\circ$ (see Fig. 7f). 

The distribution of orbital parameters and the resulting scattered line
profile is dependent upon $\beta$ because $\beta$ influences the lifetime
of the particle.  Dynamical lifetimes of particles are greater for smaller
$\beta$.  
 For Comet 2P/Encke particles with very short ($\le$5 Kyr) dynamical lifetimes
(e.g., for `2P 0.25t' run at $\beta$=0.05 and for `2P 0.5t' runs at $\beta$=0.2 
and $\beta$=0.1), 
 `velocity-elongation'  curves were shifted  in the velocity direction for up to
25  km s$^{-1}$ from the observational curve and from
the curves for larger ($>$50 Kyr) dynamical lifetimes 
(e.g., for `2P 0.25t' runs at $\beta$=0.01 and for `2P 0.5t' runs at $\beta$=0.05 
and $\beta$=0.01).
Such shift can be explained  based on studies of plots of eccentricities
versus semi-major axes. 
At smaller $\beta$, particles migrate more slowly into the Sun 
and interact with planets for a longer time than for larger $\beta$, and therefore
they
exhibit a wider range of eccentricities, even though they have the same
origin. 
`Velocity-elongation' and `eccenticity-semi-major axis' plots for such runs
are presented in astro-ph/0608141. 
As the motion is stochastic and
the number of particles in our runs is not large, there may be no strict
dependence on $\beta$. At $\beta$$\ge$0.1,
`velocity-elongation' curves for `2P 0.25t' and `2P 0.5t' runs
are higher than the observational line. 
In this case, all eccentricities are large at $a$$>$1 AU,  
plots of $a$ versus $e$ are practically the same for all particles, and 
dynamical lifetimes of all particles are close to each other and are very short ($<$5
Kyr).
For $\beta$=0.05, the maximum dynamical lifetime of Comet 2P particles
launched at aphelion (`2P 0.5t' run) was greater by a factor of 4 than that
for
`2P 0.25t' run. Therefore at $\beta$=0.05, the scatter of values of $e$ at the same
$a$ was greater for `2P 0.5t' run than for `2P 0.25t' run.
For many particles other than Comet 2P particles, 
dynamical lifetimes are several tens or several hundreds of thousands of years
and can reach tens of millions of years (Ipatov and Mather 2006).
Besides, more particles were produced by Comet 2P in its perihelion
(in this case, the plots do not differ much from the observational plot
even at large $\beta$) than in aphelion.
Therefore the zodiacal light contribution is very small for particles with
$\beta$$\ge$0.05 produced by Comet 2P at aphelion or in the middle of the
orbit.

\section{Sources of Zodiacal Dust Particles} 

\subsection{Our estimates based on observations used in previous studies}

In this subsection we show that some observations 
used in the previous publications for estimates of the fraction of cometary
particles in the zodiacal cloud 
does not contradict to the values of
the fraction greater than those presented in Table 1.
Based on these observations and results of our calculations, we also
study fractions for other dust sources.

{\it Number density at $R$$>$$3$ AU.}
First we consider the fractions that fit the observations
of the number density $n(R)$. 
For particles originating inside Jupiter's orbit, 
$n(R)$  decreases quickly with distance $R$ from the Sun 
at $R$$>$3 AU (Ipatov and Mather 2006). 
For 39P runs and $\beta$$\ge$0.002,
$n(R)$  was greater at $R$=3 AU than at $R$$\sim$5-10 AU, 
and it was greater for smaller $R$ at $R$$<$3 AU.
Therefore the fraction of particles originating
beyond Jupiter's orbit among overall particles at $R$=3 AU
can be considerable (and even dominant) in order to fit
{\it Pioneer}'s  10 and 11 observations, which showed 
(Hummes 1980; Gr\"un 1994) that $n(R)$$\approx$const 
at $R$$\sim$3-18 AU and masses $\sim$10$^{-9}$-10$^{-8}$ g 
($d$$\sim$10 $\mu$m and $\beta$$\sim$0.05). 
Otherwise one must explain
why particles  migrated from 7 to 3 AU disappear somewhere.
A considerable fraction of the particles originating
beyond Jupiter's orbit 
is also in agreement with our studies of the Doppler shift 
of Mg I line (Section 5.2) and the below studies of
the distribution of number density between 1 and 3 AU.
The number density of
trans-Neptunian particles at $R$$\sim$5-10 AU is smaller by a
factor of several than that at $R$$\sim$20-45 AU.
Therefore in order to fit  $n(R)$$\approx$const, 
 the fraction of trans-Neptunian particles at $R$$\sim$5-10 AU must be
smaller by a factor of several than the fraction of particles produced
by comets at such $R$, and we can expect that at  $d$$\sim$10 $\mu$m
the fraction of trans-Neptunian dust among zodiacal particles
is smaller by a factor of several than the fraction of cometary particles
originated beyond Jupiter's orbit and probably doesn't exceed 0.1.
We also consider that the fraction of trans-Neptunian particles
colliding with the Earth is probably greater than the estimate 
(0.01-0.02) by Moro-Martin and Malhotra (2003).

$<$ Place for Figure 8 
$>$
\centerline{}

{\it Number density at $R$$<$$3$ AU.}   
In Fig. 8a-c 
we present the values of $\alpha$ in $n(R)$$
\propto$$R^{-\alpha}$ for $R$ equal to 0.3 and 1 AU (a), at $R$=0.8 and
$R$=1.2 AU (b), and at $R$ equal to 1 and 3 AU (c). 
Collisions can shorten lifetimes of particles
and change the distributions of particles over  
$\beta$ and orbital elements,  
but we do not think that 
for the considered range of distances between 0.3 and 3 AU, 
collisions cause variations of $\alpha$
greater than the differences between values of $\alpha$
obtained at close values of $\beta$ (see also discussion
of the role of collisions in Section 5.2).
Below in this section we compare 
the values of $\alpha$ presented in Fig. 8   
with the values of $\alpha$ deduced from 
observations of the actual zodiacal cloud. The
micrometeoroid flux ($10^{-12}$ g - $10^{-9}$ g, i.e., at $d$$<$5 $\mu$m
for $\rho$=2.5 g cm$^{-3}$, or at $\beta$$\ge$0.1) measured on board {\it 
Helios 1}
during 1975 is compatible with $n(R)$$\propto$$R^{-1.3}$ at
$R$ between 0.3 and 1 AU
(Gr\"un et al. 1977; Leinert et al. 1981).
Observations by Earth's satellite {\it IRAS} yielded 
$n(R)$$\propto$$R^{-1.1}$ (Reach 1991).
{\it Pioneer 10} observations between the
Earth's orbit and the asteroid belt yielded
$n(R)$$\propto$$R^{-1.5}$ for particles of mass $\sim$$10^{-9}$ g (Hanner et
al. 1976; Reach 1992). Based on integrations of the motion of
 1-5 $\mu$m particles, at 0.5$\le$$R$$\le$1.5 AU Ozernoy (2001) obtained that
$\alpha$=1.5-1.7 for the cometary dust,
$\alpha$=1.4 for the trans-Neptunian dust, and $\alpha$=1.0 for the
asteroidal dust.

   In our models at 0.3$\le$$R$$\le$1 AU and
0.001$\le$$\beta$$\le$0.2, all values of
$\alpha$ exceed 1.9 for Comet 2P particles and are smaller than 1.1 for
asteroidal particles (Fig. 8a). 
At $\beta$$\ge$0.02,
the values of $\alpha$ for particles originating
from other considered comets were less than 1.5, but were mainly greater
than those
for asteroidal particles and in some runs exceeded 1.3.
For two-component dust cloud model, $\alpha$=1.3 can be produced
if we consider 86\% of particles
with $\alpha$=1.1 and 14\% of particles with $\alpha$=2.
It means that the fraction of Comet 2P particles needed to fit the {\it 
Helios} observations is probably less than 0.15.
Let us use other data to estimate this fraction.
Dynamical lifetimes are different
for particles of different sizes and different origin.  
 The mean value of dynamical lifetimes of 100 $\mu$m asteroidal 
and Comet 10P particles 
is about 0.5 Myr (Ipatov and Mather 2006).
According to  Fixsen and Dwek (2002), 
the total mass of the zodiacal cloud
is (2-11)$\cdot$10$^{15}$ kg. 
Using the above assumptions, we obtain the influx of zodiacal dust
to be about 1300-7000 kg s$^{-1}$, though this is quite crude estimate.
Lisse et al. (2004) obtained that the dust mass loss
rate was between 70-280 kg s$^{-1}$ for Comet 2P/Encke at $R$=1.17 AU.
If we take the mean values for the above two intervals, we obtain the
fraction of Comet 2P/Encke particles among overall
zodiacal particles to be about 0.04.
The production 
of dust by Comet 2P/Encke at perihelion (at 0.33 AU)
is greater than that at 1.17 AU.
Smaller bodies moving in Encke-type orbits also produce dust.
Therefore the fraction of particles similar to Comet 2P particles can exceed 0.1.
Dynamical lifetimes of {\it lp} and {\it ht} particles are small
at $\beta$$>$0.02, and so the fraction of such particles
in the overall population is small
at $d$$<$20 $\mu$m. Observations of the number density
were made for small particles, and they
doesn't allow one to make conclusions on the fractions of {\it lp}
or {\it ht} particles at $\beta$$\le$0.01.

   At $\beta$$\ge$0.1 and 0.8$\le$$R$$\le$1.2 AU,
the mean value of $\alpha$ for all points in Fig. 8b  
was a little smaller
than 1.5. For cometary dust, $\alpha$ was mainly greater than for
asteroidal dust; this difference was greater at $\beta$$\le$0.05 than at 
$\beta$$\ge$0.1. For $\beta$$\le$0.2, the values of 
$\alpha$ for Comet 2P particles
were greater than for other sources of dust considered.
At 1$\le$$R$$\le$3 AU for most of the dust sources,
the values of $\alpha$  were mainly greater than
the observed value equal to 1.5 (Fig. 8c).  
At 0.1$\le$$\beta$$\le$0.2, the
values of $\alpha$ for particles originating from trans-Neptunian
objects and Comet 39P/Oterma
better fit the observational value of 1.5 than those for particles from
other sources (including asteroidal dust).
This is another argument that fraction of particles produced 
outside of Jupiter's orbit can be considerable.

   Based on the above conclusion that a considerable fraction 
of particles originated outside of Jupiter's orbit, 
we can infer that 
dust production rate of 
external Jupiter-family comets could be greater than the 
estimate by Landgraf et al. (2002).
If we use their estimate 
that the dust production rate of 
external
Jupiter-family comets is 80 kg s$^{-1}$ $\approx$ 2.5$\cdot$10$^9$ kg 
yr$^{-1}$
and Gr\"un's et al. (1985) estimate that the total
dust influx to the Earth is $\sim$3$\cdot$10$^7$ kg yr$^{-1}$,
then, considering that the probability of collisions of Comet 39P
particles with the Earth is $\sim$10$^{-4}$
(Ipatov and Mather 2006, 2007),
we obtain that the fraction of particles produced by
external Jupiter-family comets among particles
collided with the Earth does not exceed a few percent. 
Dust can be produced not only by evaporation and collisions, but 
also at close encounters of comets (e.g., Comet Shoemaker-Levy 9) with Jupiter.
It is possible that such encounters can considerably enhance
the production of dust at $R$$\sim$5-10 AU.

{\it Shape of microcraters}.
   In our opinion, the 
shape (diameter/depth ratio) of microcraters 
does not contradict to values of the fraction of asteroidal dust 
smaller than those obtained by 
Brownlee et al. (1993), Vedder and Mandeville (1974),
Nagel and Fechtig (1980), and Fechtig et al. (2001).
The above authors concluded that about 30\% of the dust particles
impacting upon the Lunar surface indicate material
densities of $<$ 1 g cm$^{-3}$ and therefore more
than 70\% of interplanetary particles at $R$=1 AU
are of asteroidal origin.
We consider that not all cometary particles
have such small densities because comets also include
more dense material.
Cometary particles have greater eccentricities and
inclinations at $R$=1 AU and therefore greater velocities relative
to the Moon than asteroidal particles. Hence
the probability to be captured by the Moon (or the Earth) is greater
for a typical asteroidal particle than for a typical cometary particle.
For high-eccentricity cometary particles, these probabilities can be less 
than those for asteroidal particles 
by 2 to 4 orders of magnitude (Ipatov and Mather 2006).
The ratio of the fraction of particles accreted by the Earth to the 
fraction of particles in the zodiacal cloud is different for 
different parent bodies. This difference was discussed by several 
authors (e.g., Nesvorn\'y et al. 2006).
Note that probabilities of collisions of particles with a celestial body
depend mainly not on mean values of eccentricity and
inclination, but on the fraction of particles with
small $e$ and $i$.
Therefore the difference in the probabilities can be much
greater than the difference in mean eccentricities.
To summarize the above, we conclude that less than 70\% (e.g., 30-50\%)
of asteroidal particles can also fit the observations
of the crater shape.

{\it Shape of the zodiacal cloud}. 
For our {\it tn} runs, $i_z$ was
greater than for {\it ast}, 2P, and  10P runs,
it was about the same as for 39P runs,
and it was much smaller than for {\it lp} and {\it ht} runs.
Our studies of $i_z$  indicate that it may be possible to find such
combinations of fractions of particles originating from
different comets that fit the observations of brightness vs.
latitude even without trans-Neptunian particles.
Therefore both 1/3 (suggested by Gorkavyi et al. 2000a and Ozernoy 2001) 
and even 0 for the fraction of trans-Neptunian
particles can fit the {\it COBE} observations of brightness vs. latitude.
In our runs at all $R$, including
$R$$<$1 AU, the characteristic inclinations $i_c$
of particles originating from Jupiter-family comets were
often greater than 7$^{\circ}$, and $i_c$$>$7$^{\circ}$
for some sizes of asteroidal particles (Fig. 7).
Therefore 
the  fraction of particles produced by asteroids 
and Jupiter-family comets among the optical dust cross-section 
seen in the ecliptic at 1 AU that 
fit the {\it Clementine} observations inside the orbit of Venus
can be greater than the value of 45\% obtained by Hahn et al. (2002).
Hahn et al. considered three sources:
dust from asteroids and Jupiter-family comets with the characteristic 
inclination of about 7$^{\circ}$, dust from
Halley-type comets having $i$$\sim$33$^\circ$, and an isotopic cloud
of dust from Oort Cloud comets. 

\subsection{Estimates based on the WHAM observations}

  Comparison  
of the `velocity-elongation' plots and of the mean width of the Mg~I line 
obtained at the WHAM observations
with the plots and the width 
based on our models 
provide evidence of 
a considerable fraction of cometary particles 
in zodiacal dust, but it 
does not contradict $>$30\% of asteroidal dust needed to
explain formation of dust bands.
In the future we plan to explore the fractions of particles of different
origin in the overall dust population based on various observations and
taking into account a model for the size distribution of particles. Here
we present estimates based on a much simpler, two-component zodiacal dust
cloud that fits the observations of a velocity
amplitude $v_a$.  For example, with
$v_a$=9 km s$^{-1}$ for asteroidal dust (or Comet 10P particles)
and at $v_a$=14 km s$^{-1}$ for Comet 2P particles,
the fraction $f_{ast10P}$ of asteroidal dust plus cometary particles similar
to Comet 10P particles would have to be 0.4.
If all of the  high-eccentricity cometary particles in the 
zodiacal cloud were from long-period comets
($v_a$=33 km s$^{-1}$), then $f_{ast10P}$=0.88.
Therefore for the above two-component models, we have $f_{ast10P}$$\sim$0.4-0.9,
with 1-$f_{ast10P}$ of brightness of the zodiacal cloud
due to particles produced by high-eccentricity ($e$$>$0.8) comets.
The contribution of {\it lp} particles to the zodiacal light cannot be
large because their inclinations are large and {\it IRAS} observations 
showed
(Liou et al. 1995) that most of the zodiacal light is due to particles with
inclinations $i$$<$30$^\circ$.
Also {\it lp} and {\it  ht} particles alone cannot provide constant
number density at $R$$\sim$3-18 AU.
At $\beta$$\ge$0.004, {\it lp} particles are quickly ejected from the solar
system,
so, as a rule,  among zodiacal dust we can find {\it lp} particles only 
with $d$$>$100 $\mu$m. The
contribution of {\it lp} particles to the total mass of the zodiacal cloud
is greater than
their contribution to the brightness $I$, as surface area of a particle
of diameter $d$ is proportional to $d^2$,
and its mass $M$$\propto$$d^3$, i.e., $M/I$$\propto$$d$.
Comet 2P, {\it lp}, and {\it ht} particles are  needed to compensate
for the 
small values of $v_a$ ($\sim$8-9 km s$^{-1}$) for asteroidal and Comet 10P 
particles.
Formally, the observed values of $v_a$ can be explained only by Comet 39P and
trans-Neptunian particles, without any other particles
(including asteroidal particles). Cometary particles originating
beyond Jupiter's orbit are needed to explain the observed number density
at $R$$>$5 AU, 
so the contribution of such particles 
to the zodiacal light is not small. Therefore the values of $f_{ast10P}$
can be smaller than those for the two-component models  discussed above,
but the  contribution of {\it lp} and {\it ht} particles 
(with $e_\circ$$\ge$0.975) to the
zodiacal light cannot exceed 0.1 in order to fit the observations of $v_a$.

The dynamical lifetimes of {\it lp} particles at
$\beta$$\le$0.002 (i.e., at $d$$>$200 $\mu$m) can exceed
several Myrs (i.e., can exceed mean lifetimes of asteroidal and Comet 2P 
particles). Thus
the fraction of large {\it lp} particles in the zodiacal cloud can be 
greater than
their fraction in the new particles that were produced by small bodies or
came from other regions of the solar system.
Dynamical lifetimes of dust particles are usually greater for greater $d$
(Ipatov and Mather 2006),
and some particles can be destroyed by collisions with other particles.
Therefore the mass distributions of particles produced by small bodies are
different from the mass distributions of particles located
at different distances from the Sun.

Our studies presented above do not contradict to the model
of the zodiacal cloud for which fractions of asteroidal particles,
particles originating beyond Jupiter's orbit (including
trans-Neptunian particles), and cometary particles
originating inside Jupiter's orbit are about 1/3 each,
with a possible deviation from 1/3 up to 0.1-0.2. 
As it is discussed in Introduction and Section 5.1,
 a considerable fraction of cometary particles 
among zodiacal dust is in accordance
with most of other observations. 
Our estimated fraction of particles produced by 
long-period and Halley-type comets
in zodiacal dust  does not exceed 0.1-0.15.
The same conclusion can be made for particles originating from 
Encke-type comets (with $e$$\sim$0.8-0.9).

Though our computer model is limited, the main conclusions on
the fractions of particles of different origin among zodiacal dust
are valid for a wider range of models.
Each `velocity-elongation' curve used in our present
studies of fractional contributions was obtained
for a fixed size of particles.
Our calculations showed that the difference between characteristic
velocities corresponding to shifts in the Mg~I line 
(or between mean eccentricities) for different sizes of particles was usually 
less than the difference for different sources of particles
(e.g., asteroidal, Comet 2P, and Comet 39P particles).
It means that reasonable variations of mass distributions
of zodiacal particles do not influence on our
conclusions about the fractions of asteroidal and cometary dust
among overall zodiacal particles. 
Eccentricities and inclinations of most zodiacal particles
are not small and their mean values usually do not differ much for
different relatively close values of $\beta$ (see Figs. 6-7). 
We expect that mean variations in orbital elements of the particles 
due to collisions are smaller than these elements
and these variations do not change our conclusions
about sources of zodiacal particles.
The collisional lifetimes of particles may be comparable or shorter 
than their dynamical lifetimes, and production of different particles
can be different at different distances from the Sun. 
For more accurate models,
collisional processes must be taken into account,
but the conclusions made in the present paper do not depend
on collisional evolution of particles. 

In our simulations of spectra of dust particles we did not take
into account that albedo can be different for particles of different
origin. Mean albedo of cometary particles is smaller than that of
asteroidal particles and interplanetary dust 
(typical albedo is 0.1$\pm$0.05 for interplanetary dust,
0.02-0.06 for comets, 0.14$\pm$0.1 for TNOs, 
0.03-0.09, 0.1-0.18, 0.1-0.22, and $>$0.3 for 
C-type, M-type, S-type, and E-type asteroids, respectively;
see Hahn et al. 2002, Fernandez et al. 2005, Grundy et al. 2005). 
Therefore the fraction of cometary particles
among overall particles will be greater than their contribution to
the zodiacal light, and our conclusion about a considerable fraction
of cometary dust will be only enhanced. 

There is no considerable difference in the ratio of fractions
of asteroidal/come\-tary/trans-Neptunian dust for our spectroscopic studies
at different 
$\epsilon$, as our studies
showed that the main contribution to the spectrum is from particles at a
distance less than 1 AU from the Earth
and similar vertical shifts of the  `velocity-elongation' curves
(different for different sources of dust)
from the observational curve were obtained for many values of $\epsilon$
(see Figs. 3-4). 
Considerable difference in the fractions 
at different $\epsilon$ may occur if we consider 
particles in different
parts of the solar system, separated by several AU.
  In the future we plan to consider mass distributions of particles.
The contribution of particles of different sizes can
depend on $\epsilon$ (but not considerably).
It is caused, for example, by the result that more massive
particles spend more time in the near-Earth ring (Ipatov et al. 2004).
Therefore the size distribution of particles at $\epsilon$ close
to 90$^{\circ}$ and 270$^{\circ}$
can differ from that for other values of $\epsilon$. 

\section{Conclusions}

Our study of velocities corresponding to Doppler shifts and widths of the 
Mg~I line in the zodiacal light 
is based on the distributions of positions and velocities
of migrating dust particles originating from various solar system sources.
These distributions were obtained from our integrations of the orbital
evolution of particles produced by asteroids, comets, and trans-Neptunian
objects.
At the elongations of the fields observed by WHAM, 
the model curves of the characteristic velocity of the line vs.
the solar elongation (`velocity-elongation' curves)
were close to each other  for several 
scattering functions considered.
The differences between the curves for several sources of dust
reached its maximum at elongation between 90$^\circ$ and 120$^\circ$.
Therefore it is important for the future zodiacal light Doppler 
shifts measurements  to pay a particular attention  to 
observing at this elongation range, since this is the elongation range 
that allows to have the best discrimination between the different dust sources.

The comparison  
of `velocity-elongation' curves and 
the 
line width averaged over the elongations 
obtained at observations made by Reynolds et al. (2004) 
with the corresponding 
curves and mean widths obtained in our models shows
that asteroidal dust particles alone cannot explain these observations,
and that particles produced by comets, including high-eccentricity comets
(such as Comet 2P/Encke and long-period comets), are needed.
The conclusion that a considerable fraction of zodiacal dust 
is cometary particles
is also supported by the comparison of the variations of a number density with 
a distance from the Sun obtained 
in our models with the spacecraft observations.  

Cometary particles originating inside Jupiter's orbit and
particles produced beyond Jupiter's orbit 
(including trans-Neptunian dust particles)
can contribute to zodiacal dust about 1/3 each,
with a possible deviation from 1/3 up to 0.1-0.2.
The fraction of asteroidal dust is estimated to be $\sim$0.3-0.5.
The estimated contribution of particles produced by long-period 
and Halley-type comets
to zodiacal dust does not exceed 0.1-0.15.
The same conclusion can be made for particles originating from
 Encke-type comets (with $e$$\sim$0.8-0.9).

The velocity amplitudes of `velocity-elongation' curves 
are greater for greater mean eccentricities and inclinations,
but they depend also on distributions of particles over their orbital
elements.
The mean eccentricities of zodiacal particles
located at 1-2 AU from the Sun
that better fit the WHAM observations
are between 0.2 and 0.5, with a more probable value of about 0.3.

\section*{Acknowledgements}

     This work was supported by NASA (NAG5-12265) and by the National
Science Foundation through AST-0204973.
We are thankful to David Nesvorn\'y and another reviewer for helpful discussion.


\newpage

\begin{table}
\begin{center}
\caption{Fractions of asteroidal, cometary, and trans-Neptunian 
particles among zodiacal dust, 
constrained by different observations (see details in Introduction)
}
\begin{tabular}{|p{4cm}|p{4cm}|p{2cm}|p{2cm}|p{2cm}|}
\tableline\tableline

References &  Observations used & Fraction of asteroidal dust &
Fraction of cometary dust & Fraction of trans-Neptunian dust \\
\hline
Zook 2001 & Cratering rates from Earth- and Lunar-orbiting satellites & & 
0.75 & \\
\hline
Liou et al. 1995 & IRAS observations of the shape of zodiacal cloud &  & 
0.67-0.75 & \\
\hline
Gorkavyi et al. 2000a;
Ozernoy 2001 & COBE/DIRBE observations of brightness vs. latitude & 0.30 & 
0.36 & 0.34 \\
\hline
Grogan et al. 2001; Dermott et al. 2001; Wyatt 2005 & dust bands & $>$0.3 & 
& \\
\hline
Dermott et al. 2002 & dust bands & most & & \\
\hline
Brownlee et al. 1993;  Vedder and Mandeville 1974; Fechtig et al. 2001 & 
shape of microcraters  & $>$0.7 & & \\
\hline
Nesvorn\'y et al. 2006 & IRAS observations of dust bands & 
0.05-0.09 for Karin/Veri\-tas particles & dominated by high-speed cometary
particles& \\

\hline
Present paper & WHAM observations and observations of number density & 
0.3-0.5 & 0.4-0.7 & $\le$0.1 \\

\tableline
\end{tabular}
\end{center}
\end{table}

\newpage

\begin{table}
\begin{center}
\caption{Characteristic velocity amplitude $v_a$ of `velocity-elongation'
curves at
90$^\circ$$<$$\epsilon$$<$270$^\circ$, minimum ($v_{\min}$) and
maximum ($v_{\max}$) velocities at the above interval, mean eccentricities
($e_z$) and mean
inclinations ($i_z$) at distance from the Sun 1$\le$$R$$\le$3 AU for
particles
from different sources at intervals of $\beta$}
\begin{tabular}{lllllll}
\tableline\tableline
Source of  & $\beta$ &  $e_z$  & $i_z$ & $v_a$  & $v_{\min}$ & $v_{\max}$
\\
particles  &             &           & (deg.) & (km s$^{-1}$) &(km
s$^{-1}$) & (km s$^{-1}$) \\
\tableline
asteroids & 0.0004-0.1 & $<$0.3 & 8-13 & 9   & (-13)-(-11) & 5-7 \\
asteroids & 0.0001 & 0.2-0.3 & 7-14 & 13   & -15 & 11 \\
10P        & 0.0001-0.2 & 0.2-0.6 & 7-13 & 8   & (-11)-(-9) & 3-9 \\

39P        & 0.01-0.2   & 0.2-0.4 & 10-14 & 8-9 & (-13)-(-9) & 4-8 \\
39P        & 0.0001-0.004 & 0.3-0.8 & 15-24 & 11-17 & (-20)-(-10) & 8-16 \\
2P         & 0.01-0.1  & 0.5-0.8 & 7-23 & 13 & (-12)-(-7)&12-22 \\
2P         & 0.0001-0.004  & 0.7-0.9 & 8-15 & 14 & (-16)-(-12)&15-20 \\
2P 0.25t   & 0.002-0.2  & 0.6-0.85 & 4-11 & 16  &(-14)-4& 12-40 \\
2P 0.5t   & 0.002-0.4  & 0.5-0.85 & 4-11 & 14  & (-22)-(-2)& 2-29 \\
tn       & 0.05-0.4    & 0.1-0.4 & 11-23 & 16 & (-20)-(-14)&12-18 \\
tn       & 0.002-0.01   & 0.4-0.8 & 9-33 & 12 & -16 & 8 \\
lp, $q$=0.9 AU        & 0.002 & 0.2-0.9 & 55-155 & 41 & -44 & 38 \\
lp, $q$=0.9 AU        & 0.0001-0.001 & 0.8-1. & 80-115 & 33 &
(-38)-(-32)&28-34 \\
lp, $q$=0.1 AU        & 0.0004 & 0.95-1. & 95-105 & 34 & -34 & 34 \\
ht, $q$=0.5 AU        & 0.001-0.012 & 0.45-1. & 95-135 &  37 & -35 & 40 \\

\tableline
\end{tabular}
\end{center}
\end{table}


\clearpage

FIGURE CAPTIONS:

Fig. 1.
Dependence of the intensity of light vs. its wavelength
$\lambda$
(in Angstrom) at $\beta$=0.2, $\epsilon$=180$^\circ$, in the ecliptic
plane (a) 
and at $\beta$=0.05 (exclusive for {\it lp} particles considered at $\beta$=0.002)
toward the North Ecliptic Pole (b).
Zero of $\Delta \lambda$=$\lambda$-$\lambda_\circ$ corresponds to
$\lambda$=$\lambda_\circ$=5183.62 Angstrom.
The plots for dust particles produced by asteroids, trans-Neptunian
objects, Comet 2P at perihelion, Comets 10P and 39P are denoted by `ast',
`tn', `2P', `10P', and `39P', respectively.
Data for particles originating from
long-period comets at $e_\circ$=0.995,
$q_\circ$=0.9 AU, and $i_\circ$ distributed between 0 and 180$^\circ$, are 
marked as `lp'.
Marks in Fig. (a) 
are for average intensity for observations at
174$^\circ$$\le$$\epsilon$$\le$188$^\circ$,
and marks in Fig. (b) 
are for average observations toward the North Ecliptic Pole.
Coordinates of marks for `observ/sol spectr' were obtained from
observational data
by making it have the same minimum value as the solar spectrum.
The minima of plots obtained in all calculations were made the 
same as that for the solar spectrum.

Fig. 2.
Velocities of Mg I line (at zero inclination) versus elongation
$\epsilon$ (measured eastward from the Sun) at $\beta$=0.05
for dust particles produced by asteroids (a) and Comet 2P/Encke at perihelion
(b).
Letter `c' denotes the model for which the shift of the curve of
intensity $I$ versus $\epsilon$
is calculated as a shift of centroid, and letter `m' denotes the model
for which the shift of the curve is calculated as a shift of the minimum
of the curve. The number after `m' or `c' characterizes the number of a
scattering function used.

Fig. 3.
Velocities of Mg I line (at zero inclination) versus elongation
$\epsilon$ (measured eastward from the Sun) at several values of $\beta$
(see the last number in the legend)
for particles produced by asteroids (a-b) and Comet 10P/Tempel 2 (c-d).
The line corresponds to the observations made by Reynolds et al. (2004).

Fig. 4.
Velocities of Mg I line (at zero inclination) versus elongation
$\epsilon$ (measured eastward from the Sun) at several values of $\beta$
for particles produced by Comet 2P/Encke at perihelion (a-b),
for particles launched from Comet 39P/Oterma (c-d),
for particles originating from trans-Neptunian objects (e),
test long-period comets (`lp') at $e_\circ$=0.995,
$q_\circ$=0.9 AU, and $i_\circ$ distributed
between 0 and 180$^\circ$,
test long-period comets (`lpc') at $e_\circ$=0.999,
$q_\circ$=0.1 AU, and $i_\circ$ distributed
between 0 and 180$^\circ$,
and test Halley-type comets (`ht') at $e_\circ$=0.975,
$q_\circ$=0.5 AU, and $i_\circ$ distributed
between 0 and 180$^\circ$ (f).
The line corresponds to the observations made by Reynolds et al. (2004).

Fig. 5.  
Mean value of the width of Mg I line (in the ecliptic plane) for
elongation
$\epsilon$ between 30$^\circ$ and 330$^\circ$ at several values of $\beta$
for particles originating from asteroids (ast), from Comet 2P/Encke at
perihelion (2P),
at the middle of the orbit (2P 0.25t), and at aphelion (2P 0.5t),
from Comets 10P/Tempel 2 and 39P/Oterma (10P and 39P),
from test long-period comets (lp) at $e_\circ$=0.995,
$q_\circ$=0.9 AU, and $i_\circ$ distributed between 0 and 180$^\circ$,
and from test Halley-type comets (ht) at $e_\circ$=0.975,
$q_\circ$=0.5 AU, and $i_\circ$ distributed between 0 and 180$^\circ$.
Solid line corresponds to the observational value.

Fig. 6.  
Mean eccentricity of particles at different distances from the Sun
at several values of $\beta$ (see the last number in the legend)
for particles originating from asteroids (ast), from Comet 2P/Encke at
perihelion (2P),
from Comets 10P/Tempel 2 and 39P/Oterma (10P and 
39P),
from trans-Neptunian objects (tn), from test long-period comets (lp) at
$e_\circ$=0.995,
$q_\circ$=0.9 AU, and $i_\circ$ distributed between 0 and 180$^\circ$,
from test long-period comets (lpc) at $e_\circ$=0.999,
$q_\circ$=0.1 AU, and $i_\circ$ distributed
between 0 and 180$^\circ$,
and from test Halley-type comets (ht) at
$e_\circ$=0.975,
$q_\circ$=0.5 AU, and $i_\circ$ distributed between 0 and 180$^\circ$.

Fig. 7.  
Mean orbital inclination (in degrees) of particles at different
distances from the Sun
at several values of $\beta$ (see the last number in the legend)
for particles originating from asteroids (ast), from Comet 2P/Encke at
perihelion (2P),
from Comets 10P/Tempel 2 and 39P/Oterma (10P and 
39P),
from trans-Neptunian objects (tn),
from test long-period comets (lp) at $e_\circ$=0.995,
$q_\circ$=0.9 AU, and $i_\circ$ distributed between 0 and 180$^\circ$,
from test long-period comets (lpc) at $e_\circ$=0.999,
$q_\circ$=0.1 AU, and $i_\circ$ distributed
between 0 and 180$^\circ$,
and from test Halley-type comets (ht) at $e_\circ$=0.975,
$q_\circ$=0.5 AU, and $i_\circ$ distributed between 0 and 180$^\circ$.

Fig. 8. 
Values of $\alpha$ in $n(R)$$\propto$$R^{-\alpha}$ obtained by
comparison
of the values of the number density $n(R)$ at distance $R$ from the Sun
equal to 0.3
and 1 AU (a), at $R$=0.8 and $R$=1.2 AU (b),
and at $R$ equal to 1 and 3 AU (c).
The values are presented at several values of $\beta$
for particles originating from Comets 10P/Tempel 2 and 39P/Oterma (10P and 
39P),
from trans-Neptunian objects (tn),
from test long-period comets (lp) at $e_\circ$=0.995,
$q_\circ$=0.9 AU, and $i_\circ$ distributed between 0 and 180$^\circ$,
and from test Halley-type comets (ht) at $e_\circ$=0.975,
$q_\circ$=0.5 AU, and $i_\circ$ distributed between 0 and 180$^\circ$.
Horizontal bars correspond to observations.

\newpage

\begin{figure}   

\includegraphics[width=81mm]{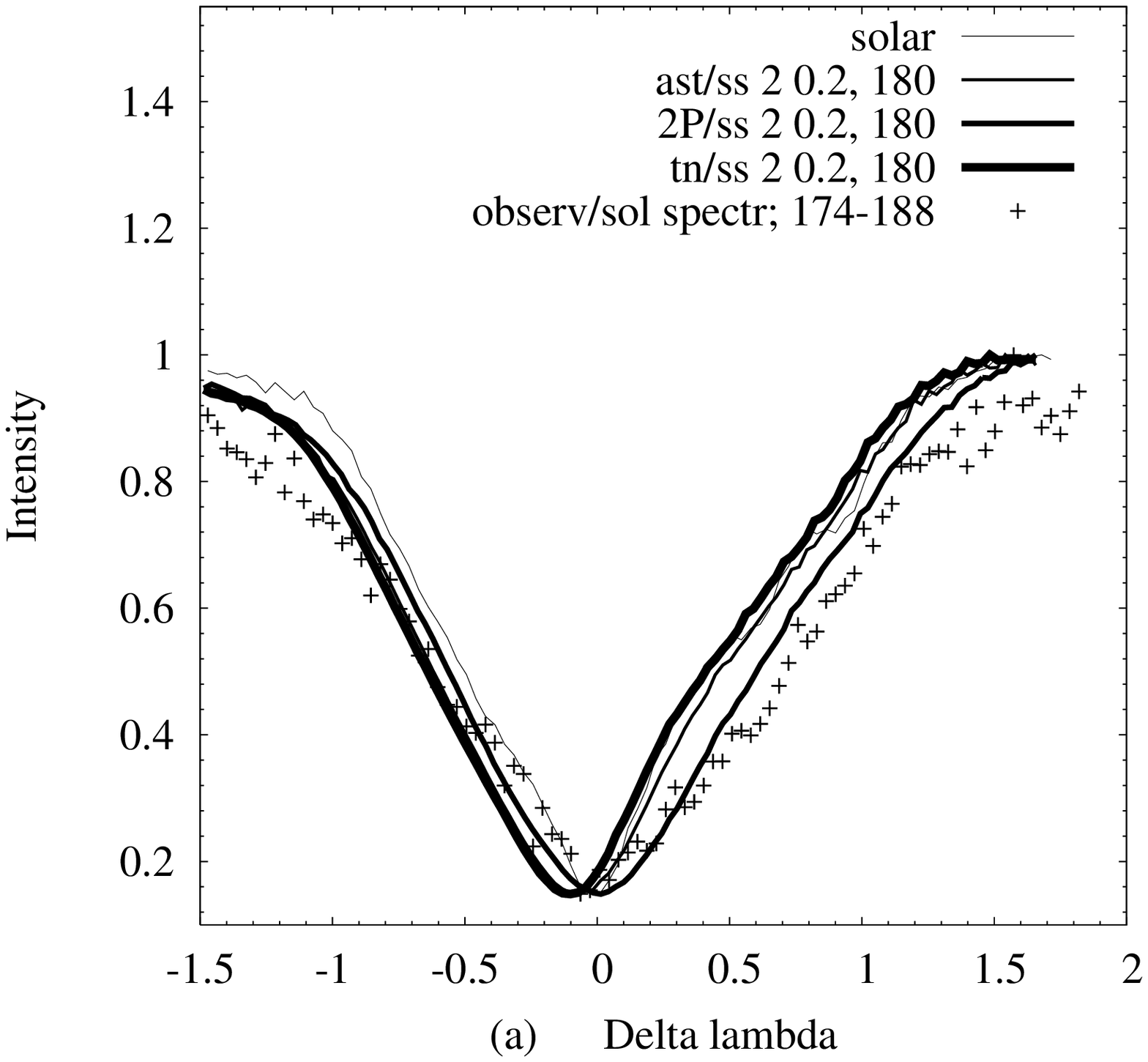}  
\includegraphics[width=81mm]{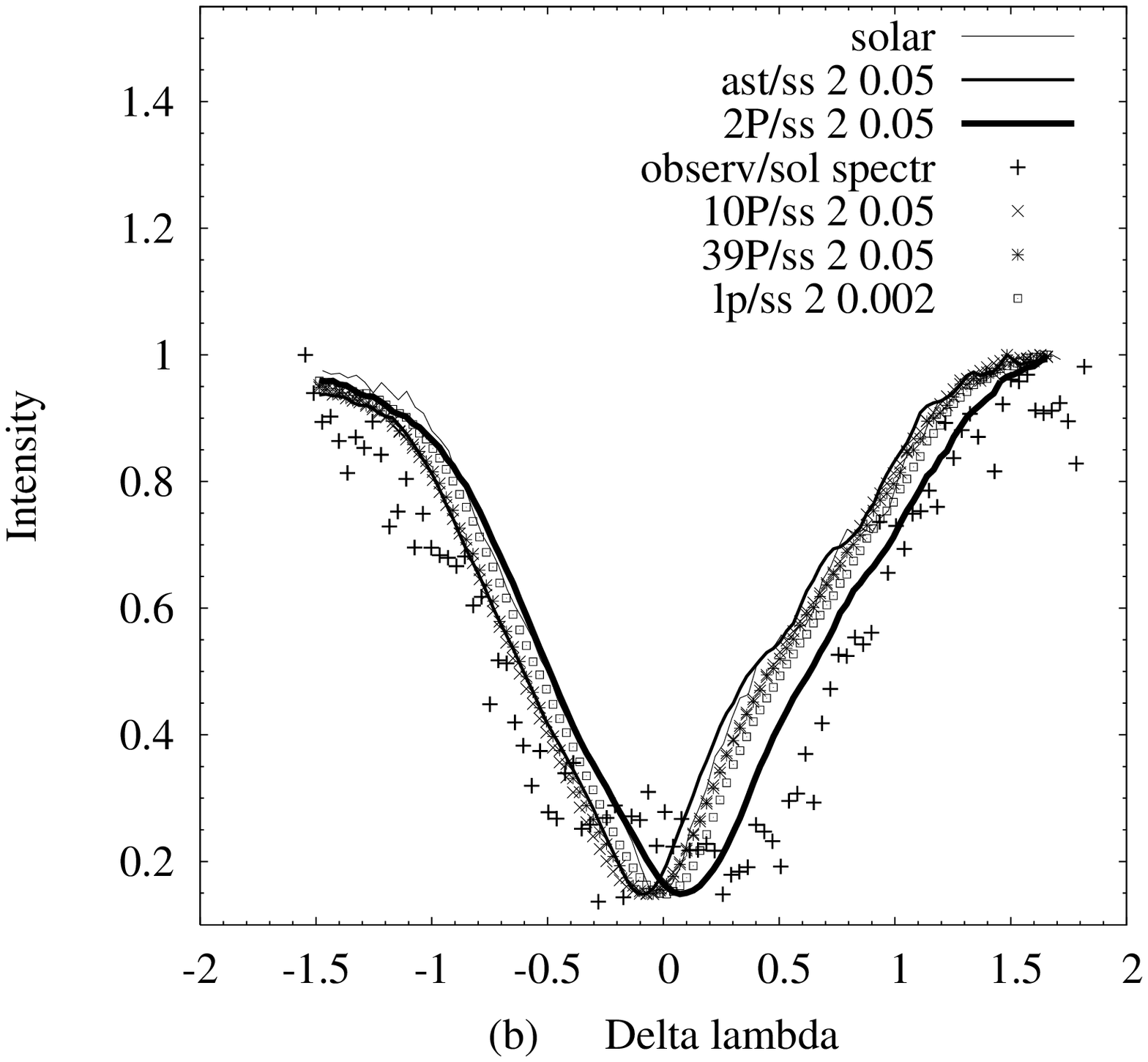}  

\caption{Ipatov et al., Zodiacal cloud...  
}
\end{figure}%

\begin{figure}   

\includegraphics[width=81mm]{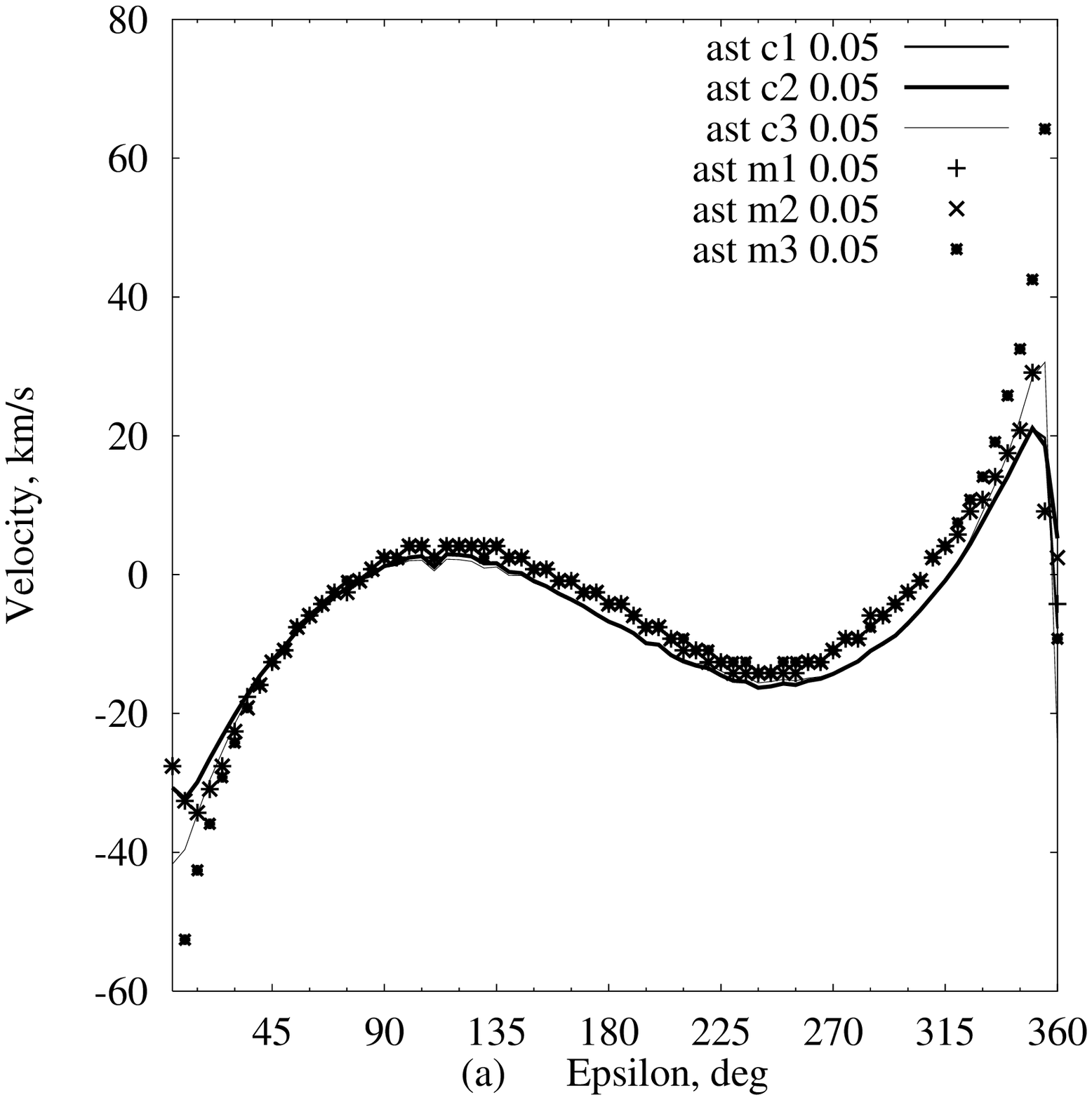}  
\includegraphics[width=81mm]{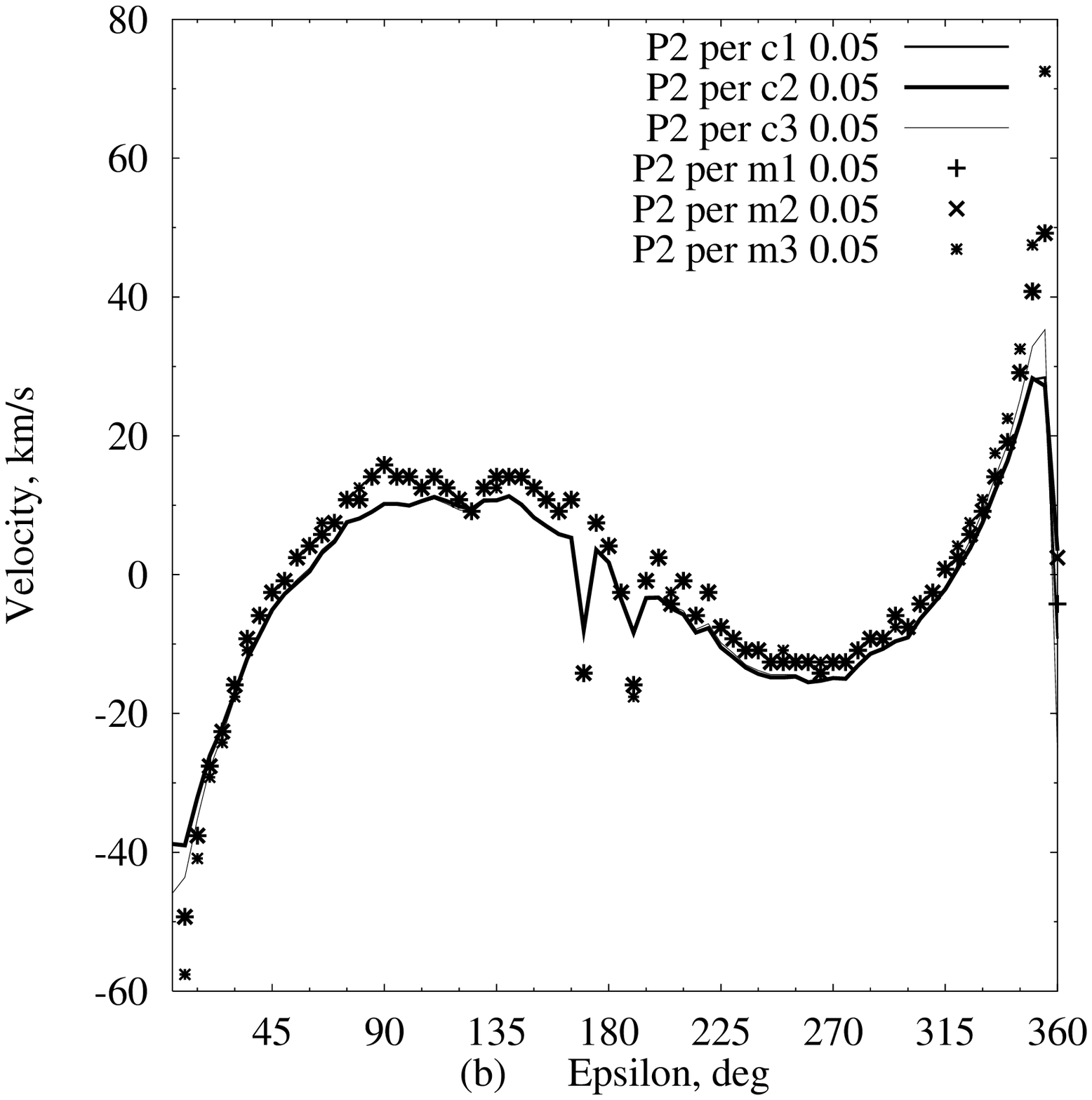}  

\caption{Ipatov et al., Zodiacal cloud...}

\end{figure}

\begin{figure}

\includegraphics[width=81mm]{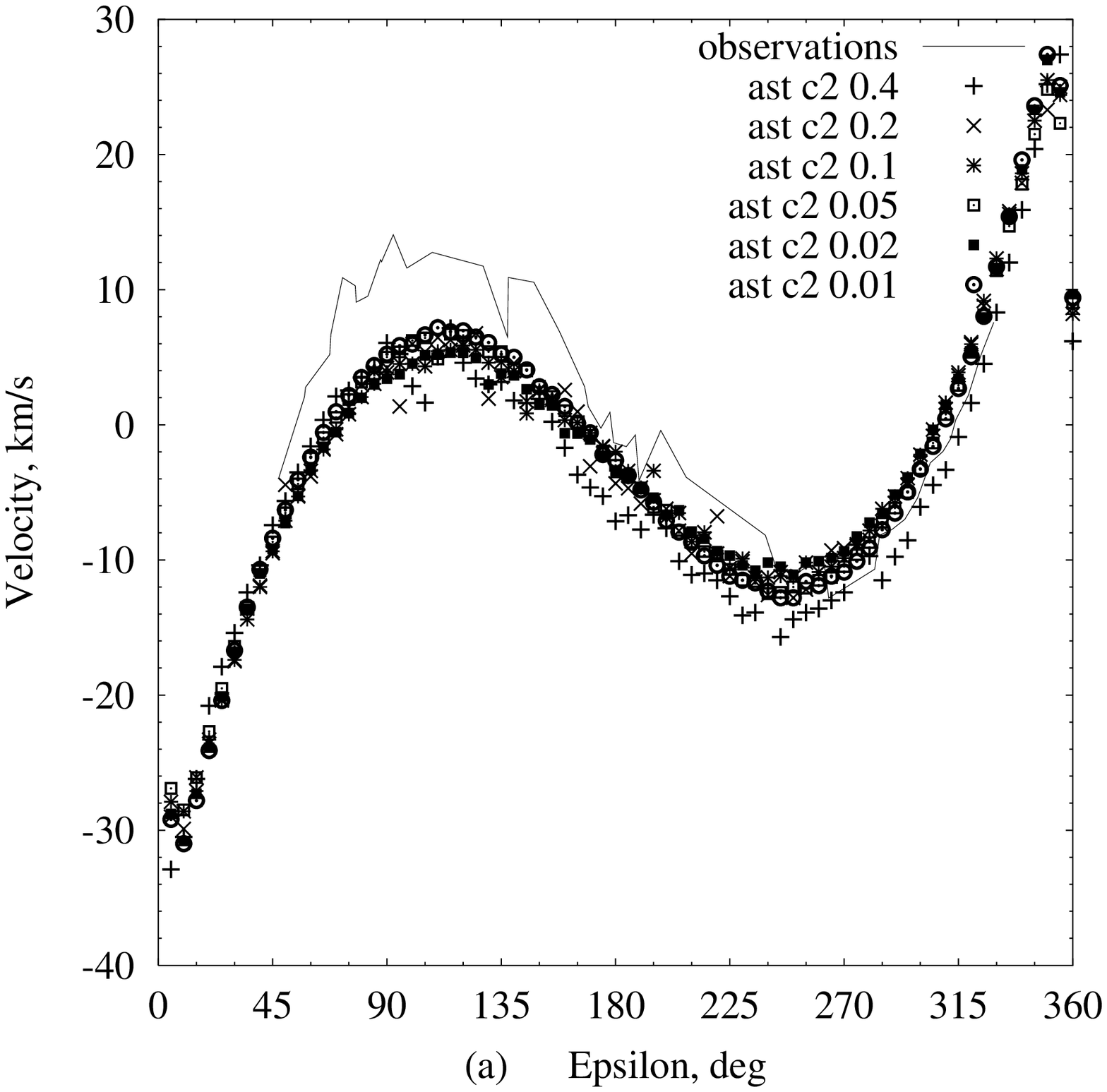}  
\includegraphics[width=81mm]{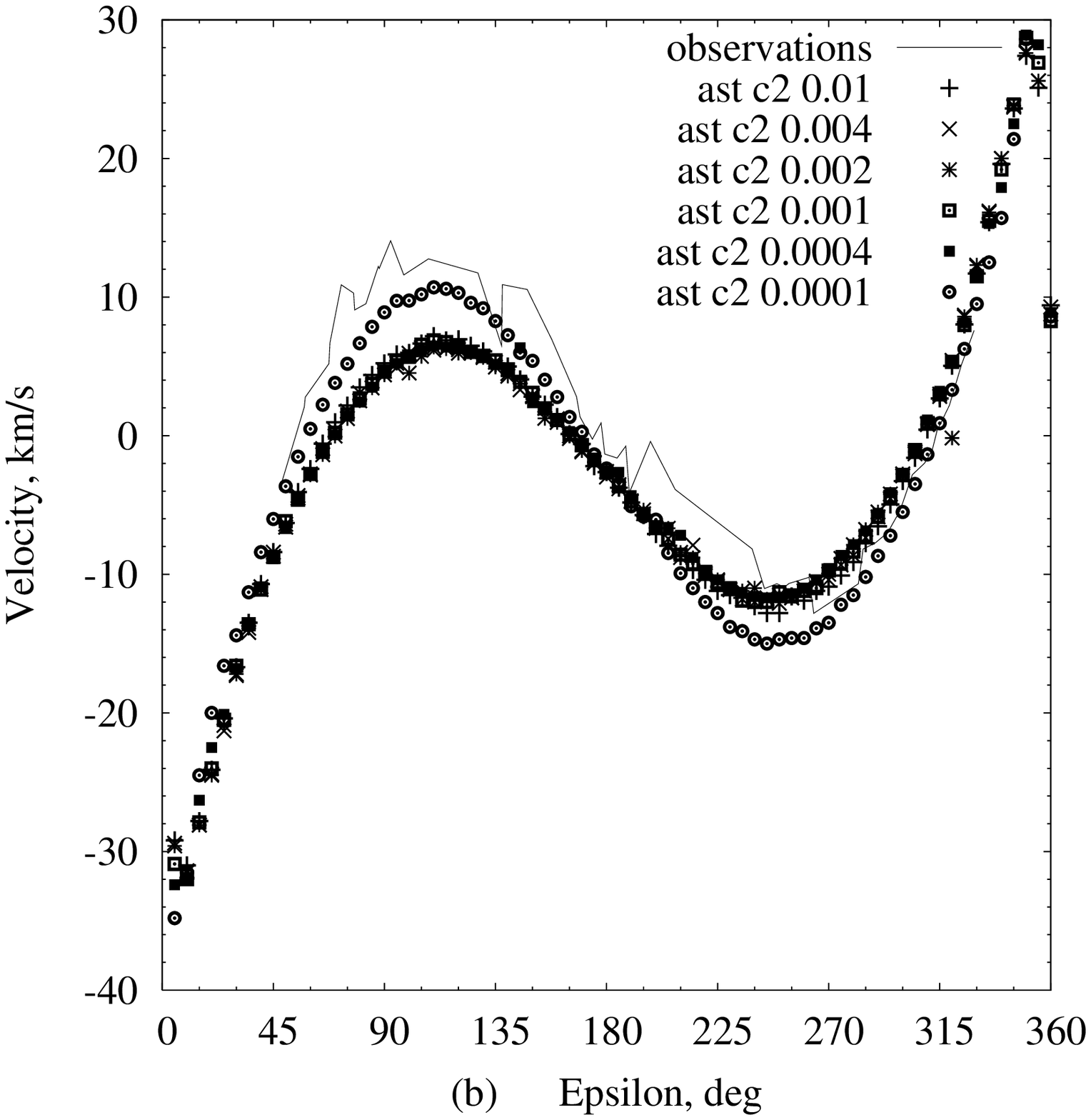}  
\includegraphics[width=81mm]{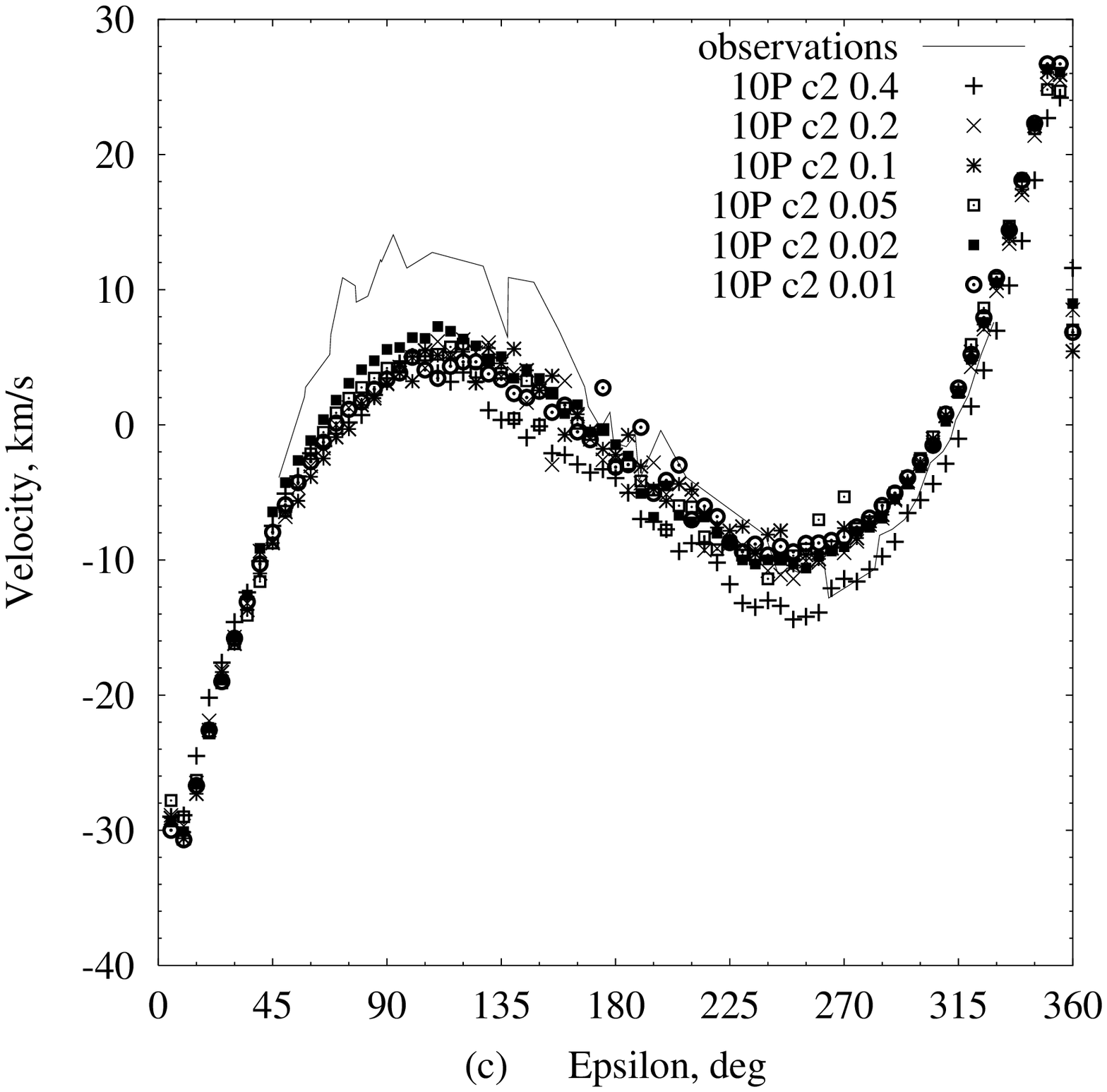}  
\includegraphics[width=81mm]{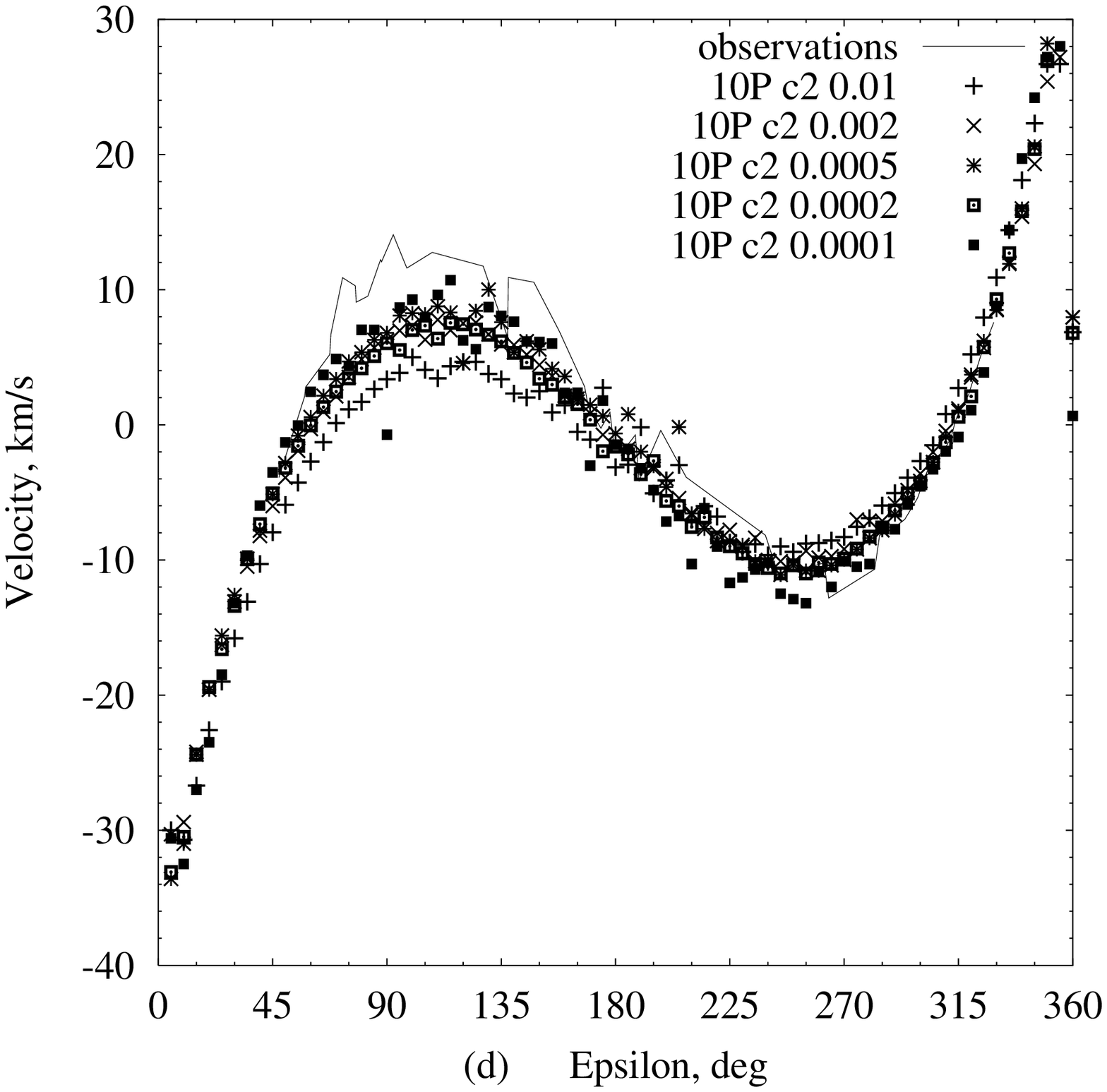}  

\caption{Ipatov et al., Zodiacal cloud...
}

\end{figure}%

\begin{figure}

\includegraphics[width=78mm]{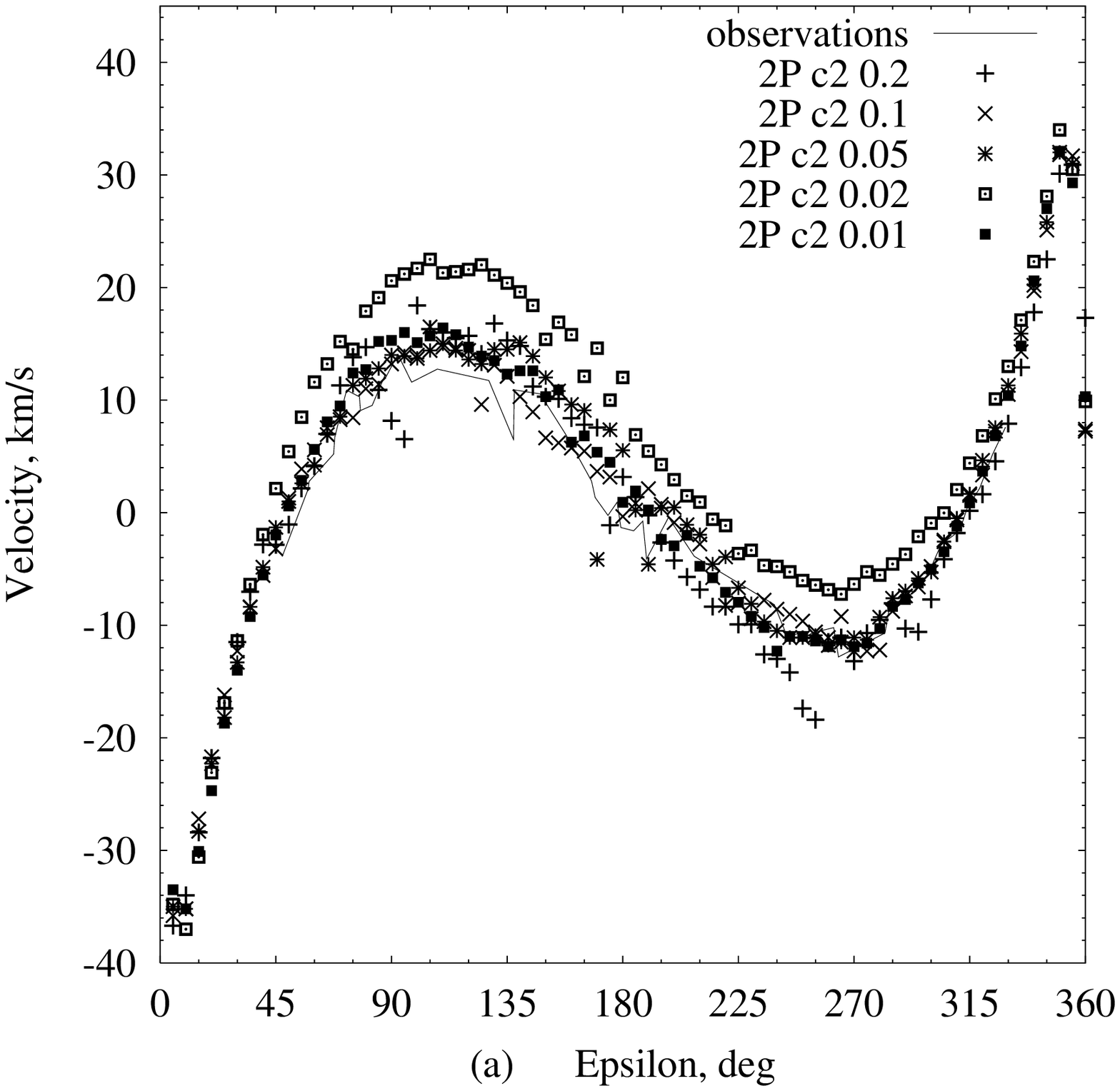} 
\includegraphics[width=78mm]{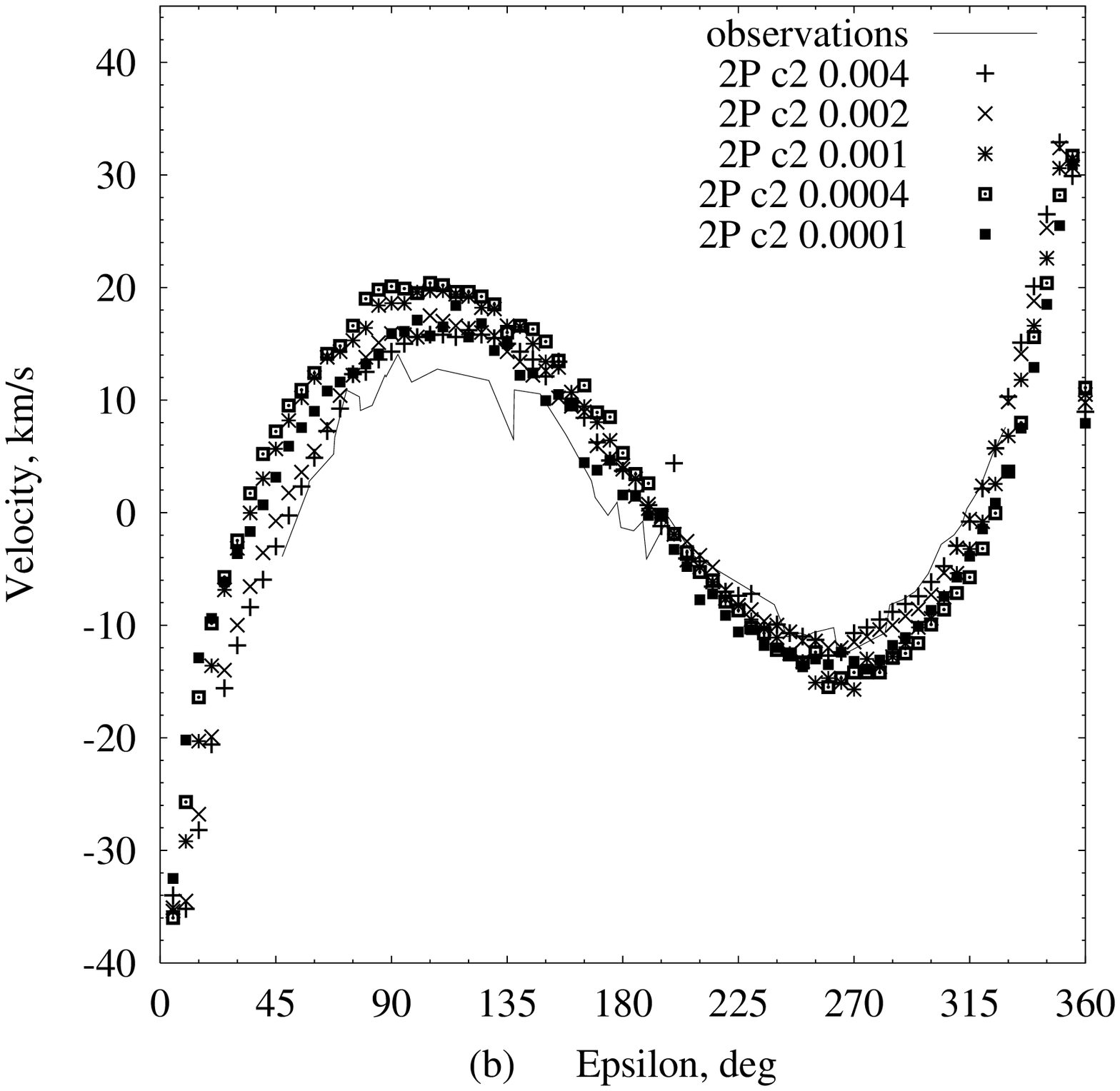}  
\includegraphics[width=81mm]{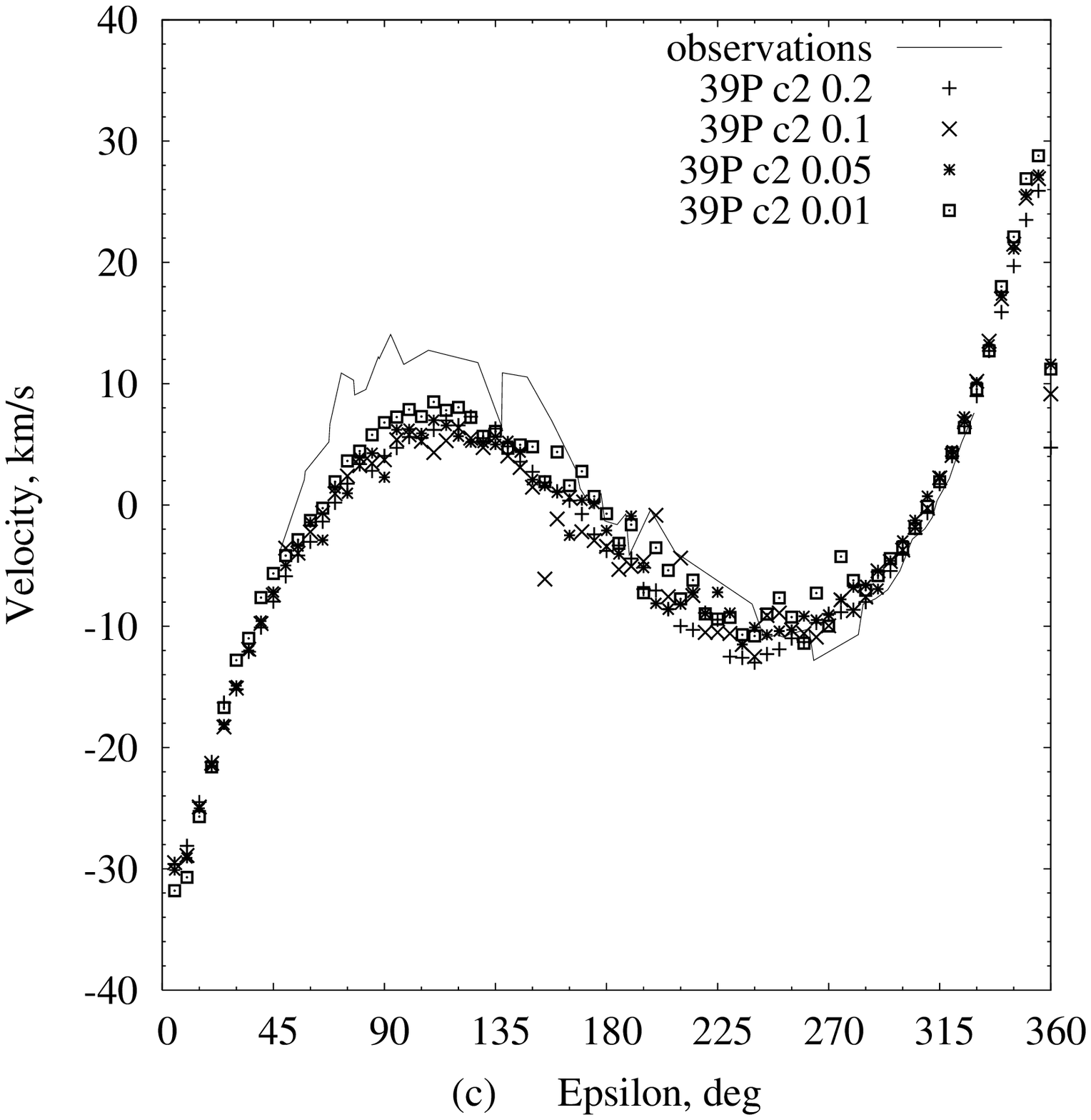} 
\includegraphics[width=81mm]{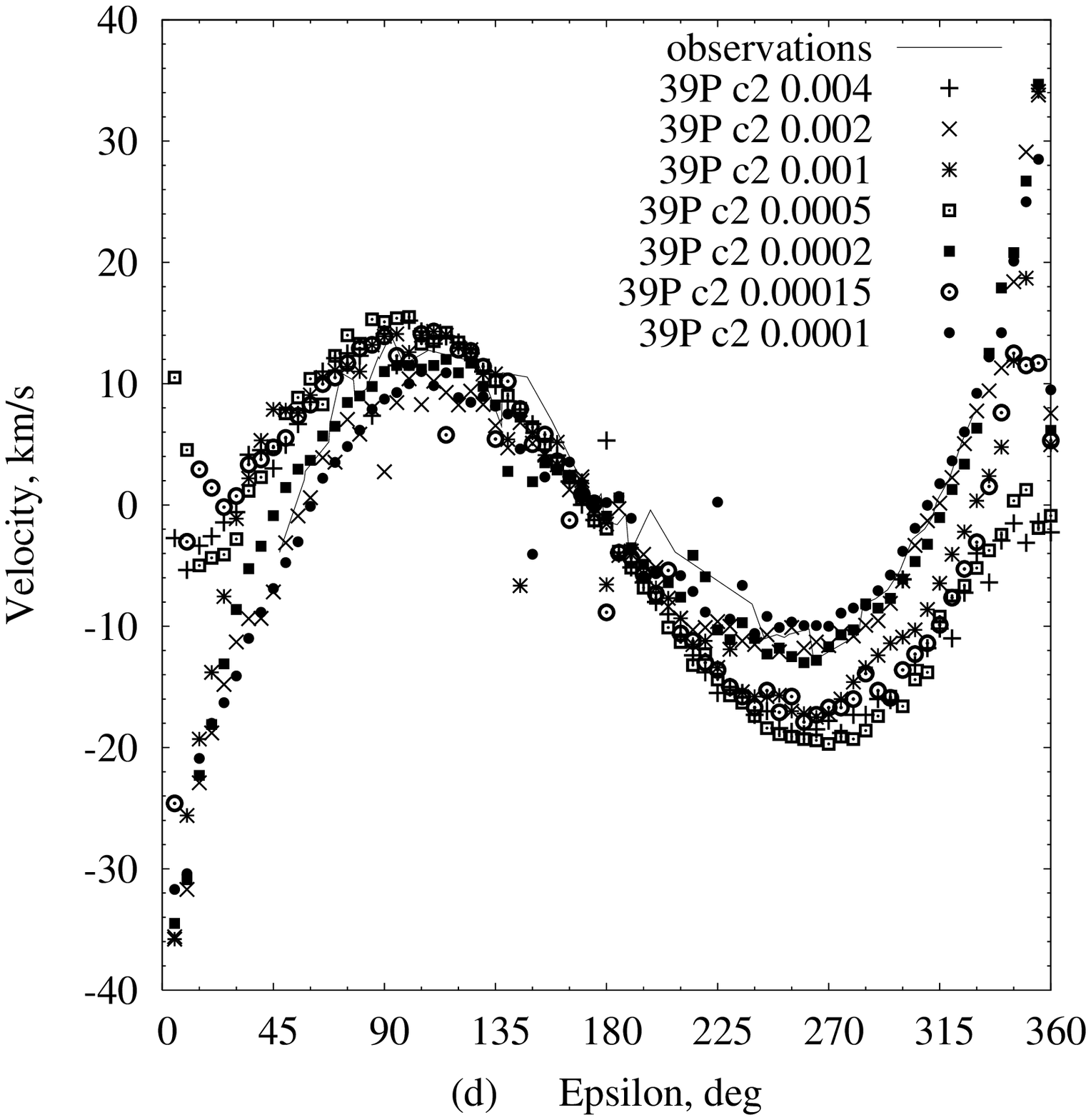} 
\includegraphics[width=81mm]{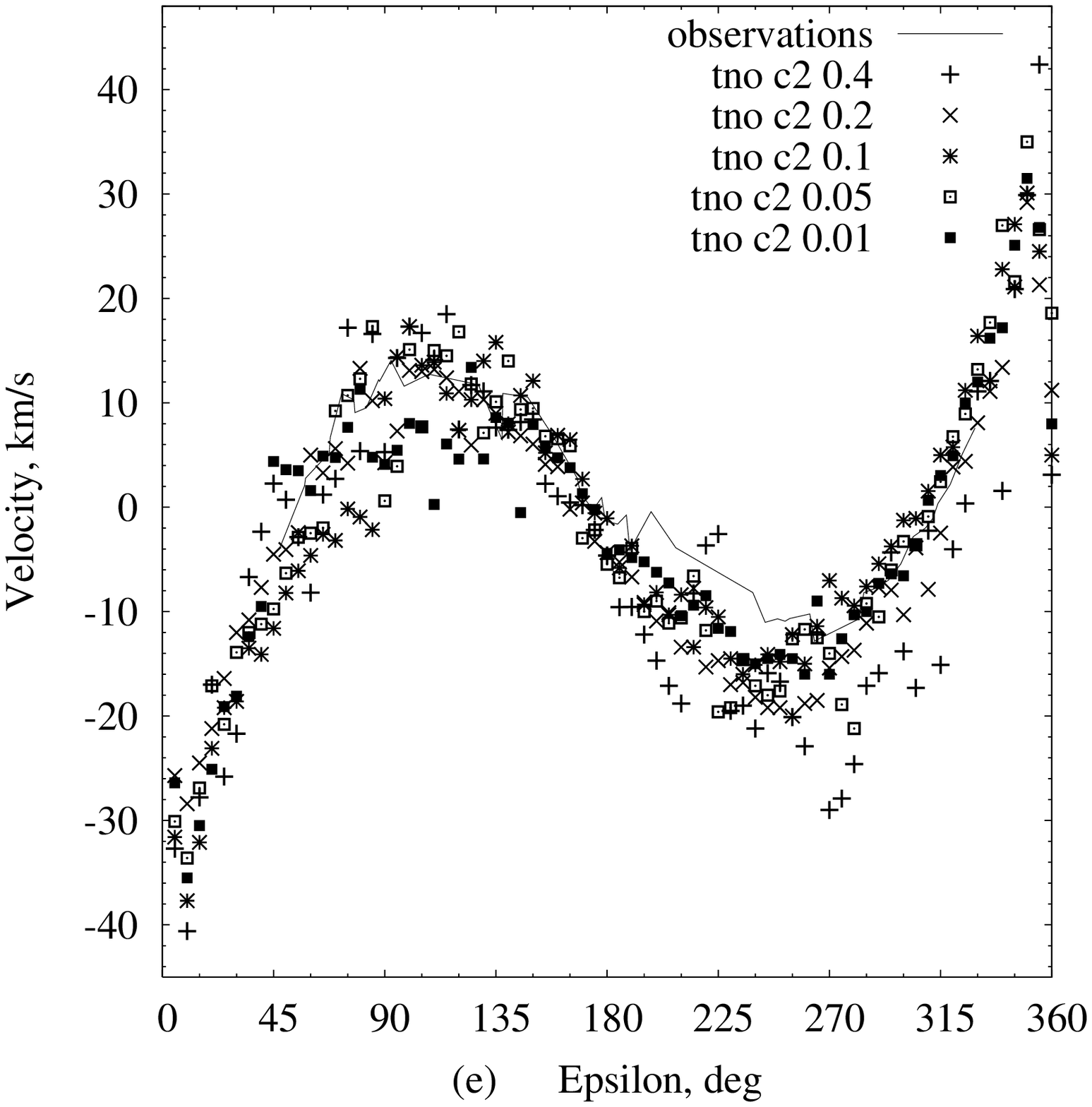} 
\includegraphics[width=81mm]{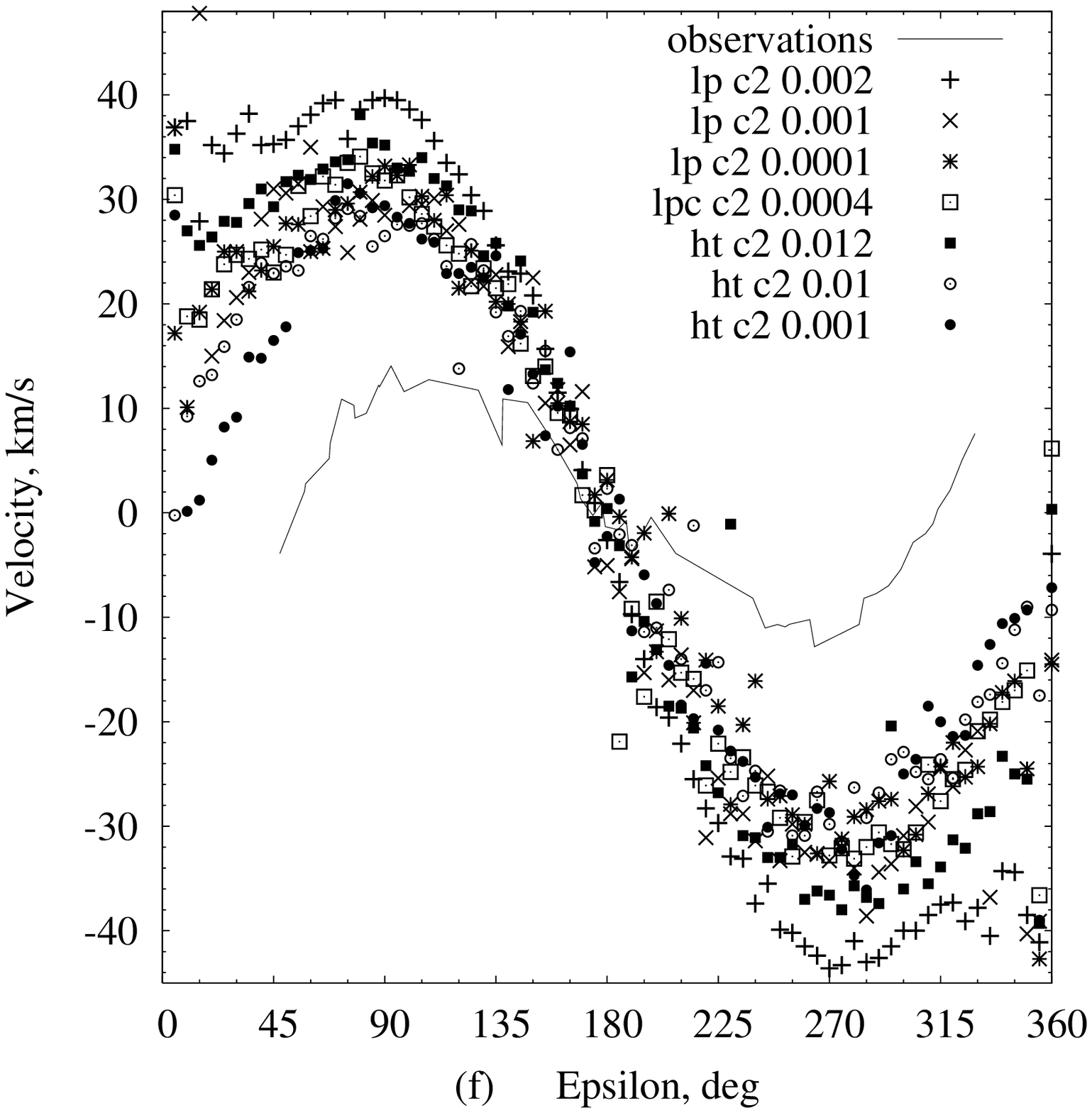} 

\caption{Ipatov et al., Zodiacal cloud...}

\end{figure}%

\begin{figure}    

\plotone{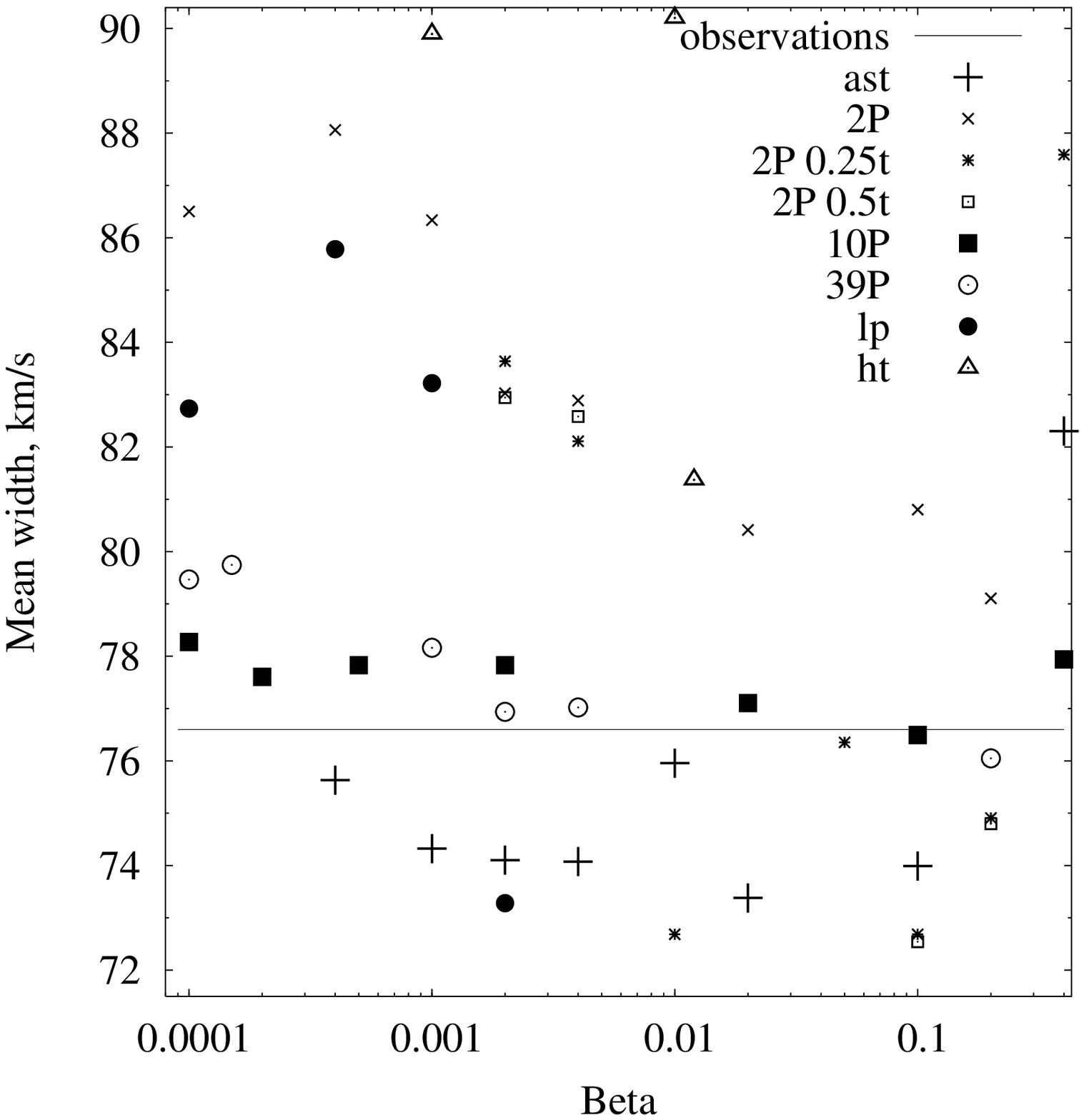} 

\caption{Ipatov et al., Zodiacal cloud...
}

\end{figure}%

\begin{figure}   

\includegraphics[width=81mm]{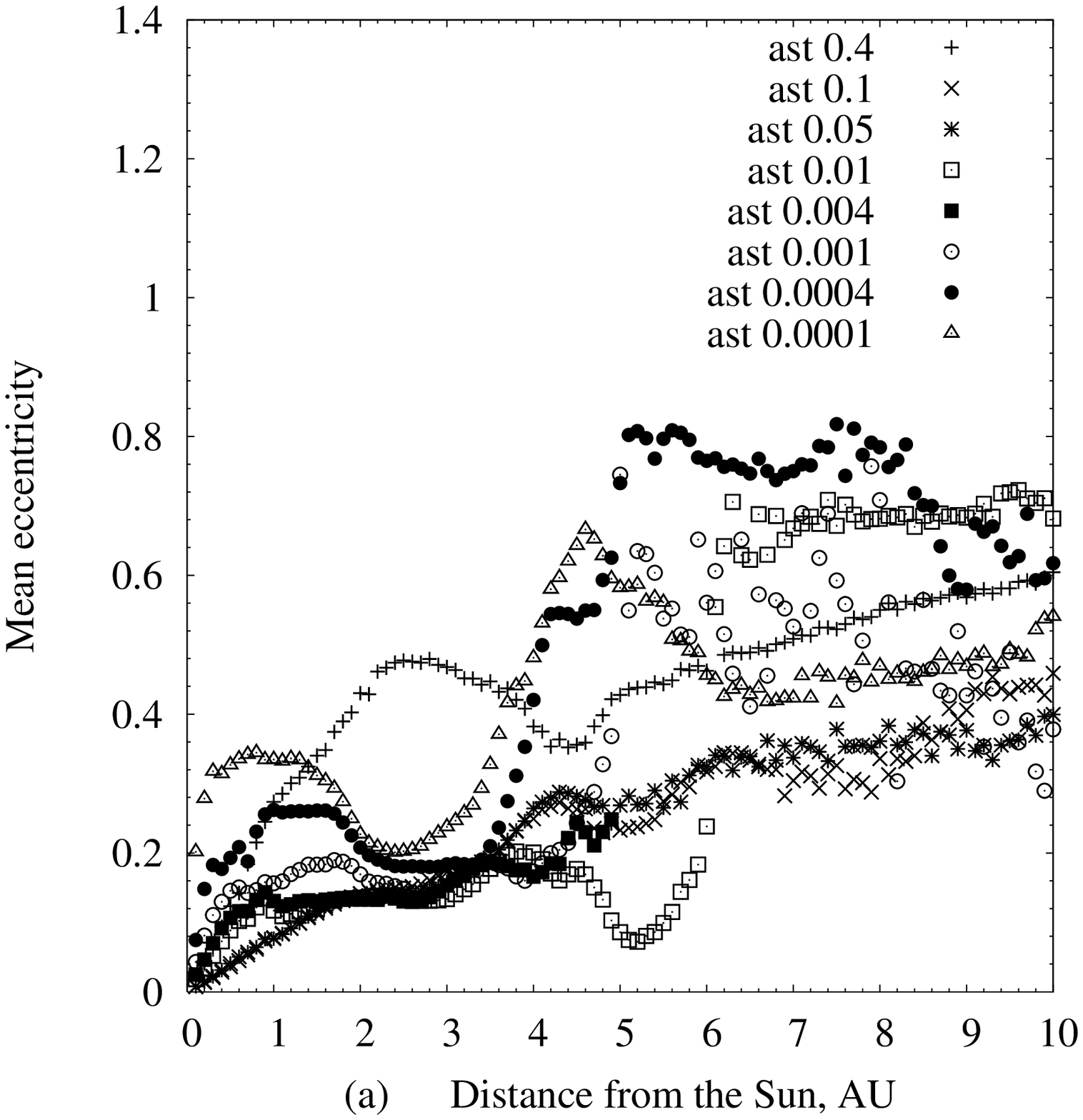} 
\includegraphics[width=81mm]{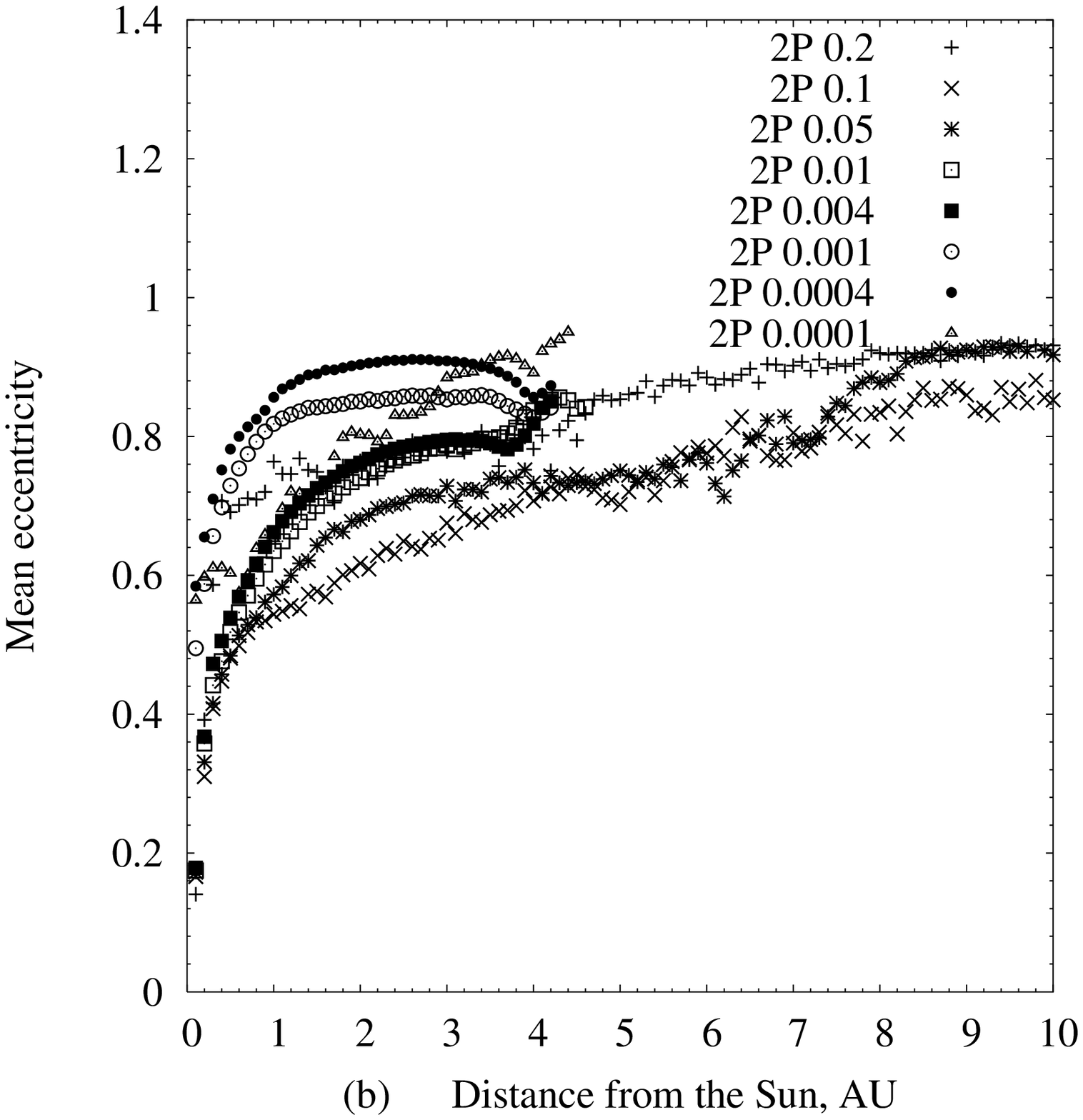} 
\includegraphics[width=81mm]{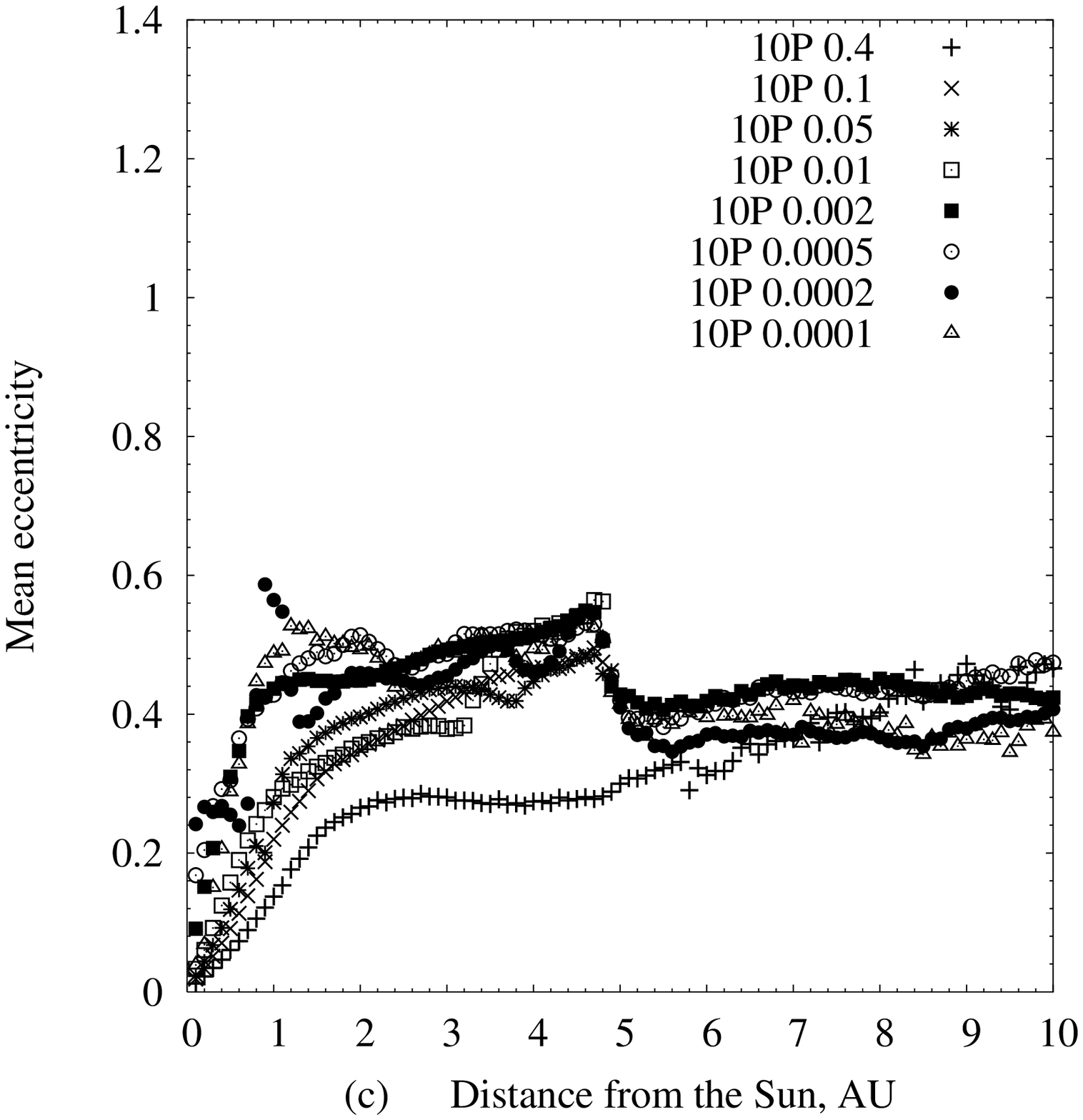}
\includegraphics[width=81mm]{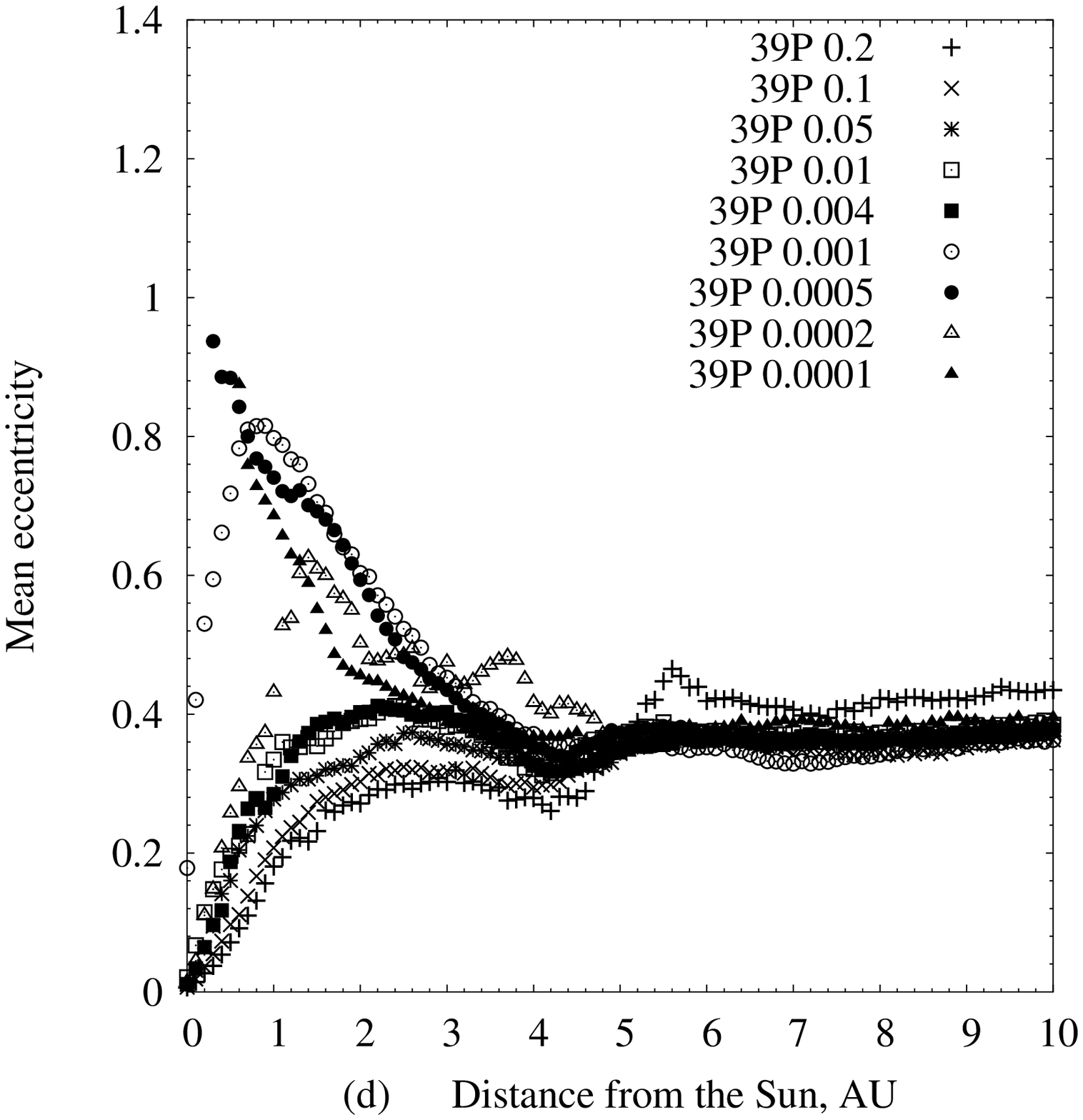} 
\includegraphics[width=81mm]{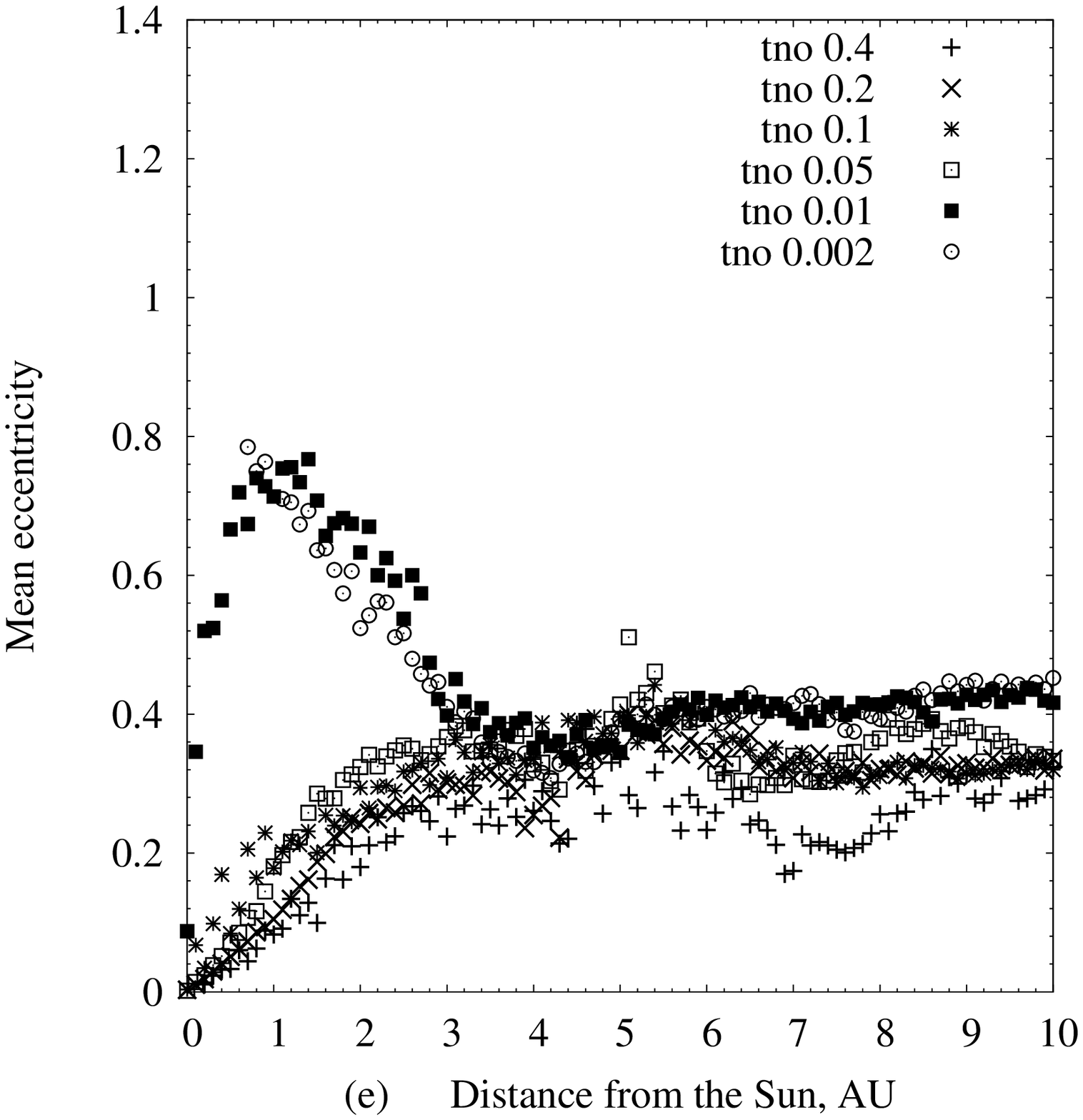} 
\includegraphics[width=81mm]{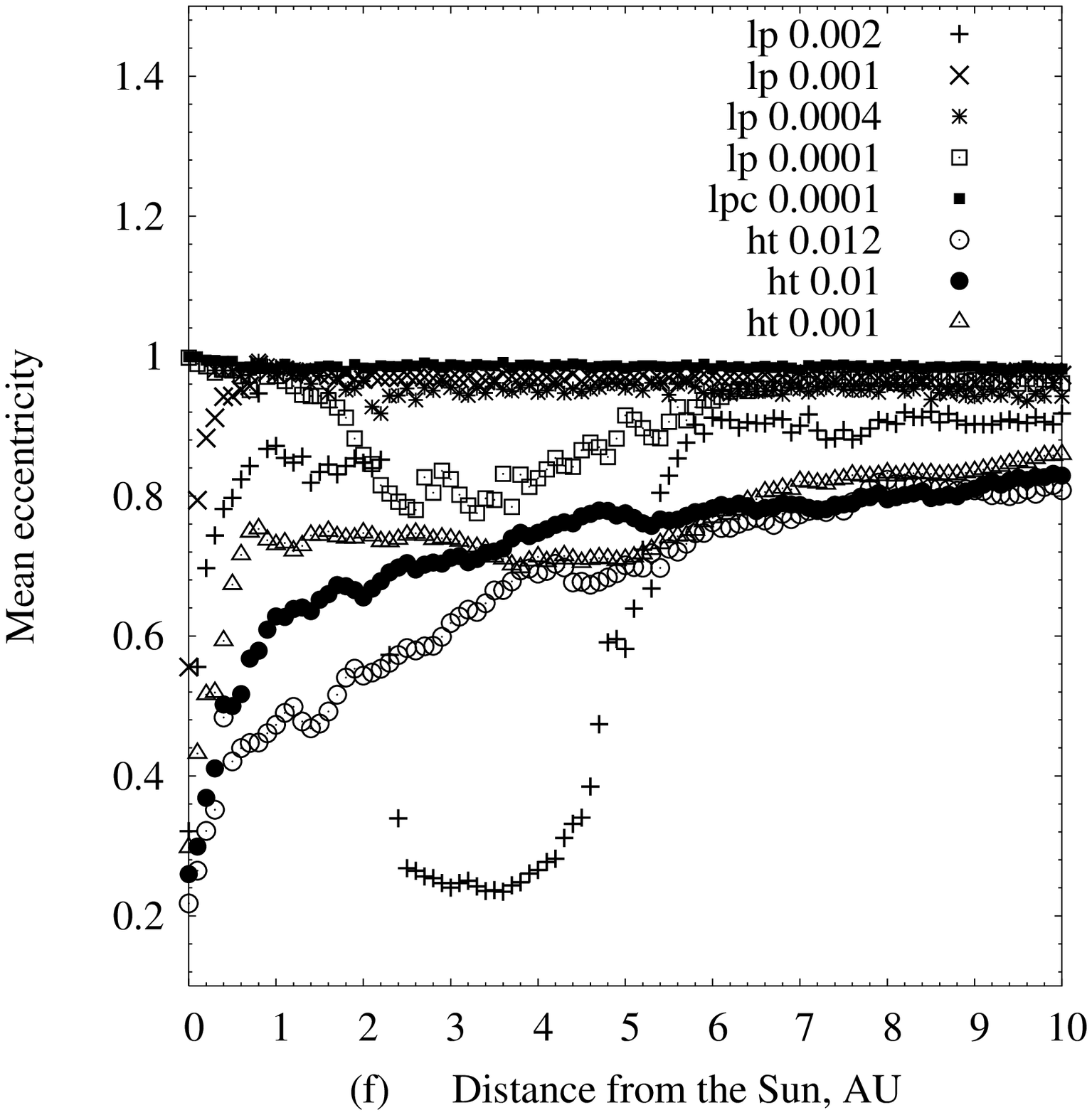} 

\caption{Ipatov et al., Zodiacal cloud...}
\end{figure}%

\begin{figure}    

\includegraphics[width=81mm]{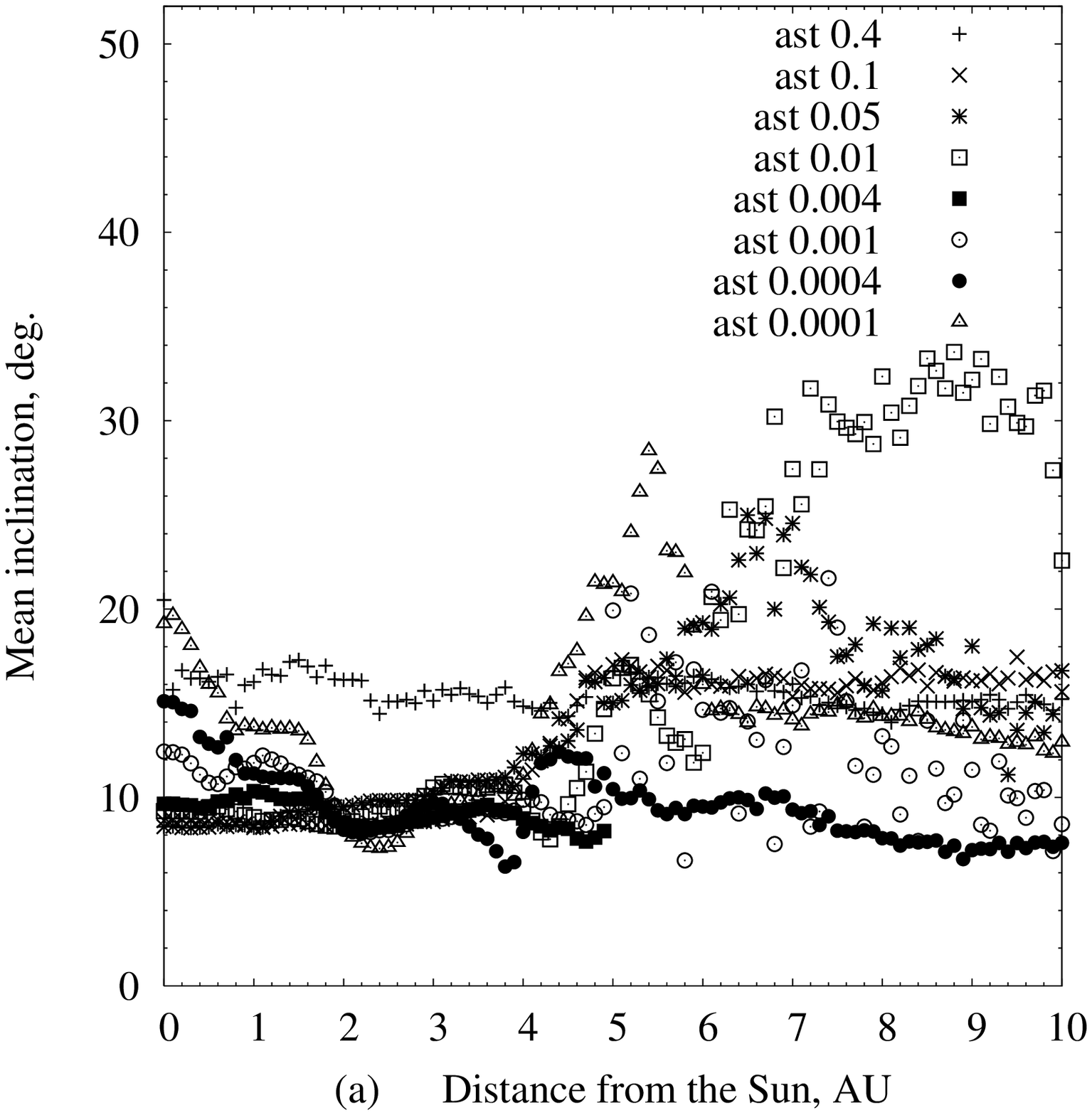} 
\includegraphics[width=81mm]{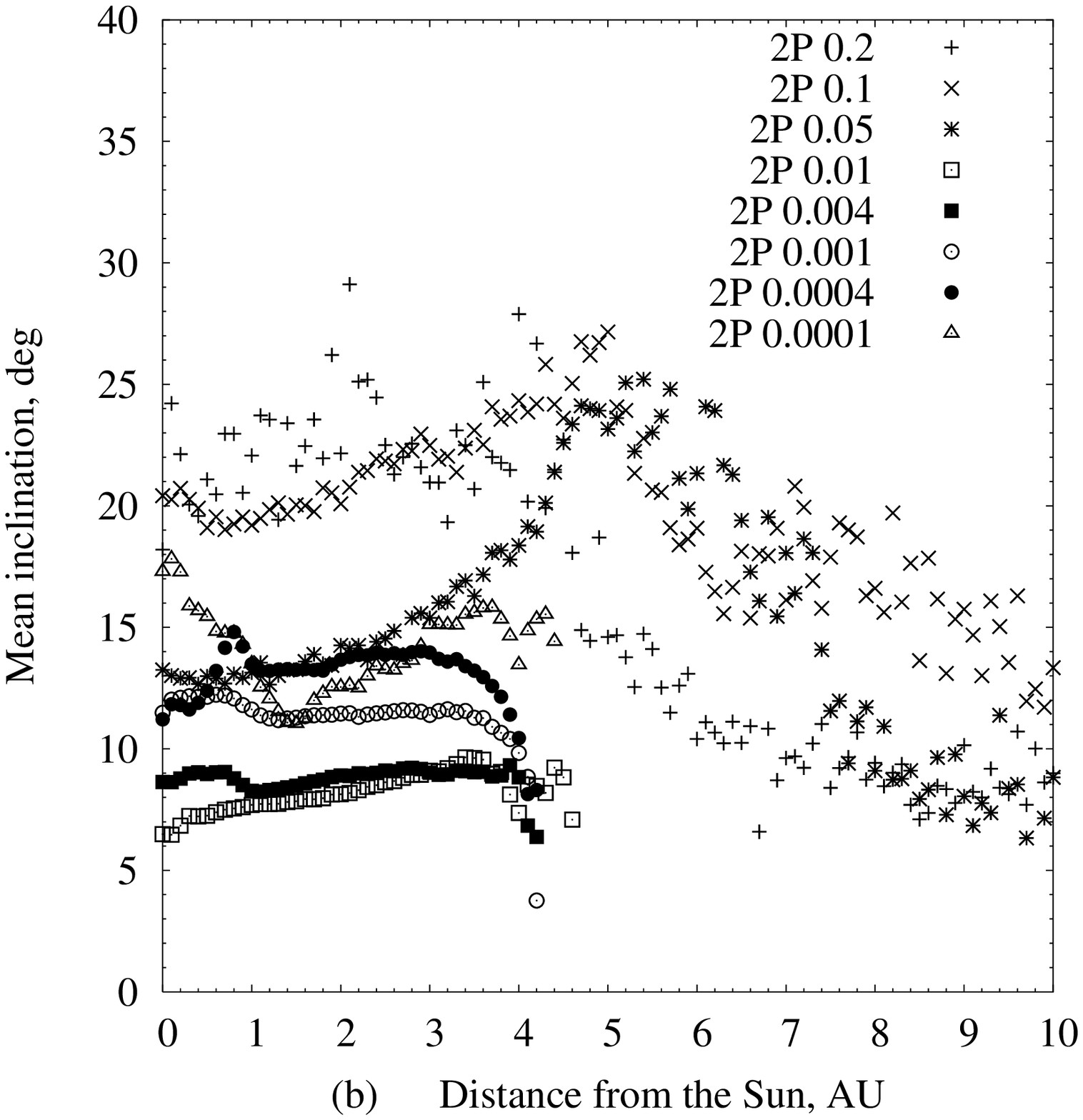} 
\includegraphics[width=81mm]{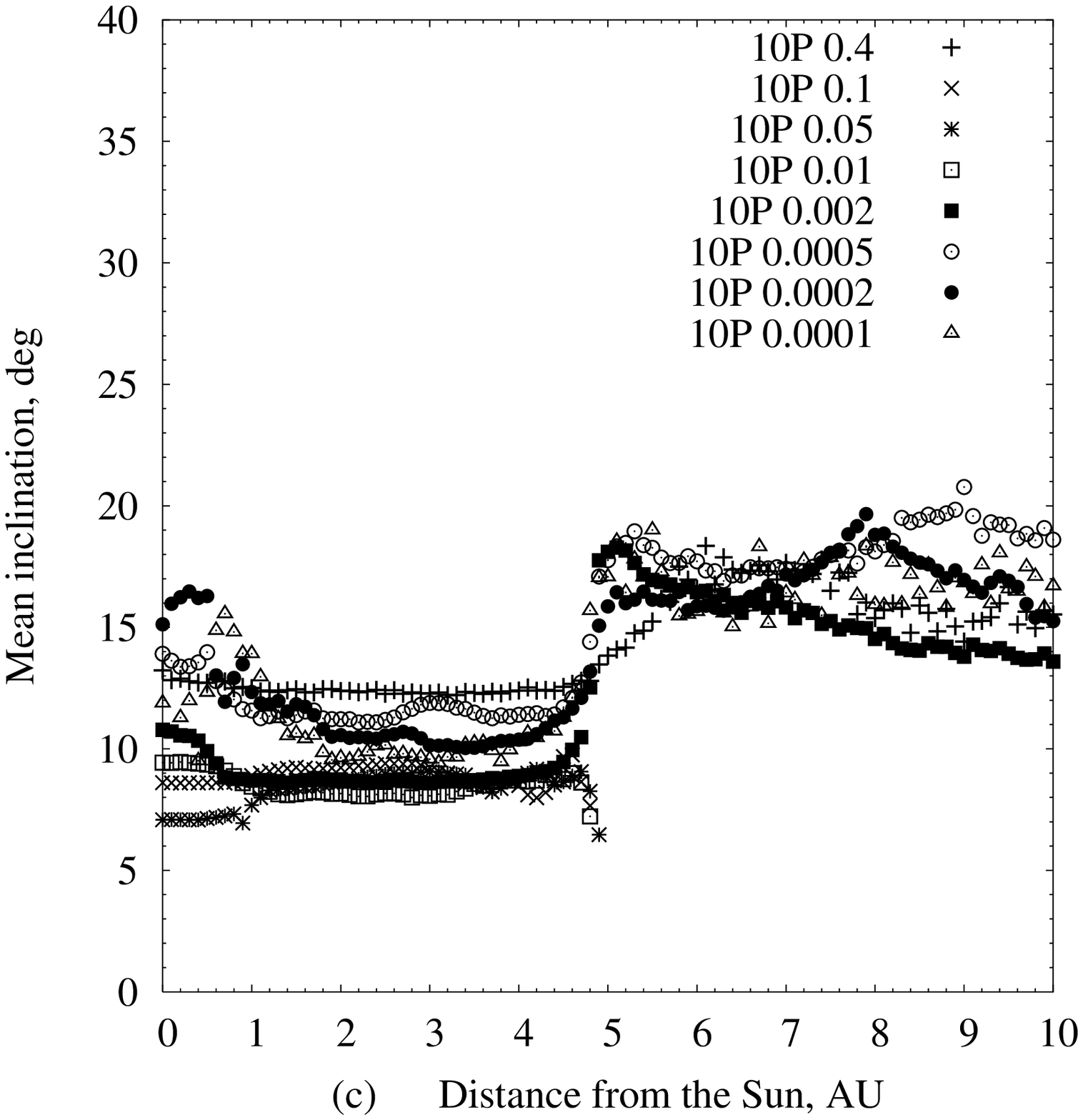} 
\includegraphics[width=81mm]{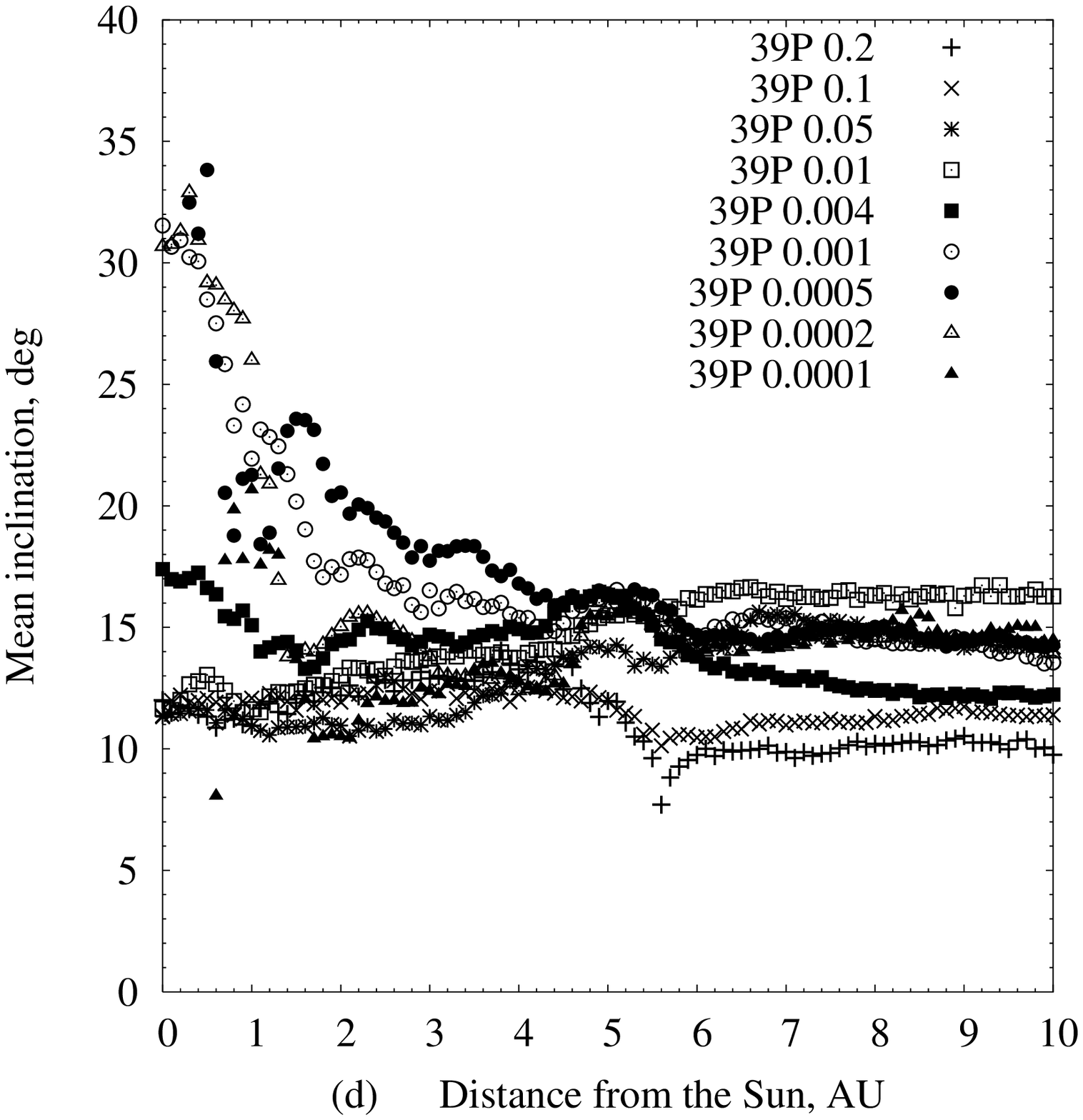} 
\includegraphics[width=81mm]{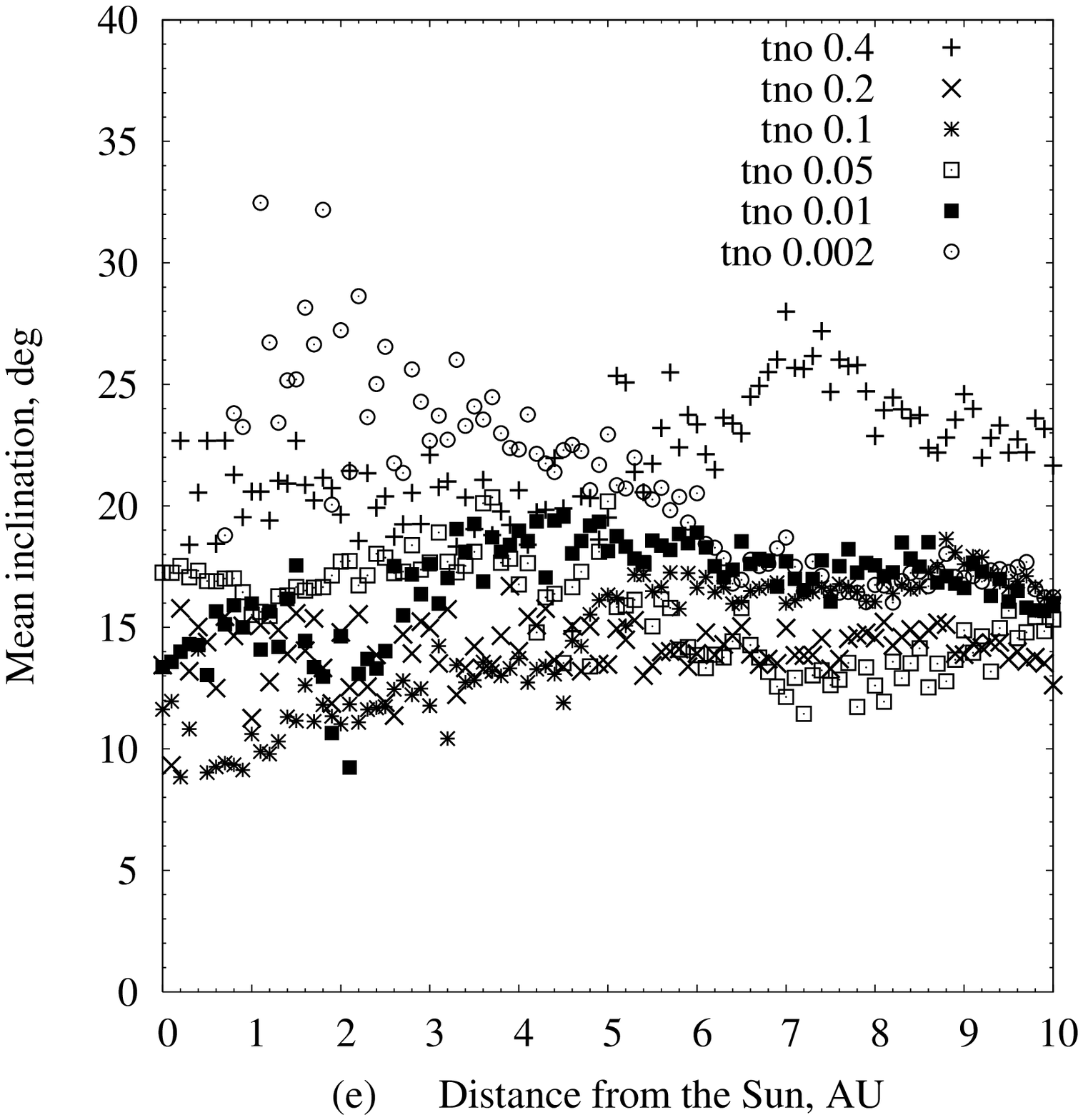} 
\includegraphics[width=81mm]{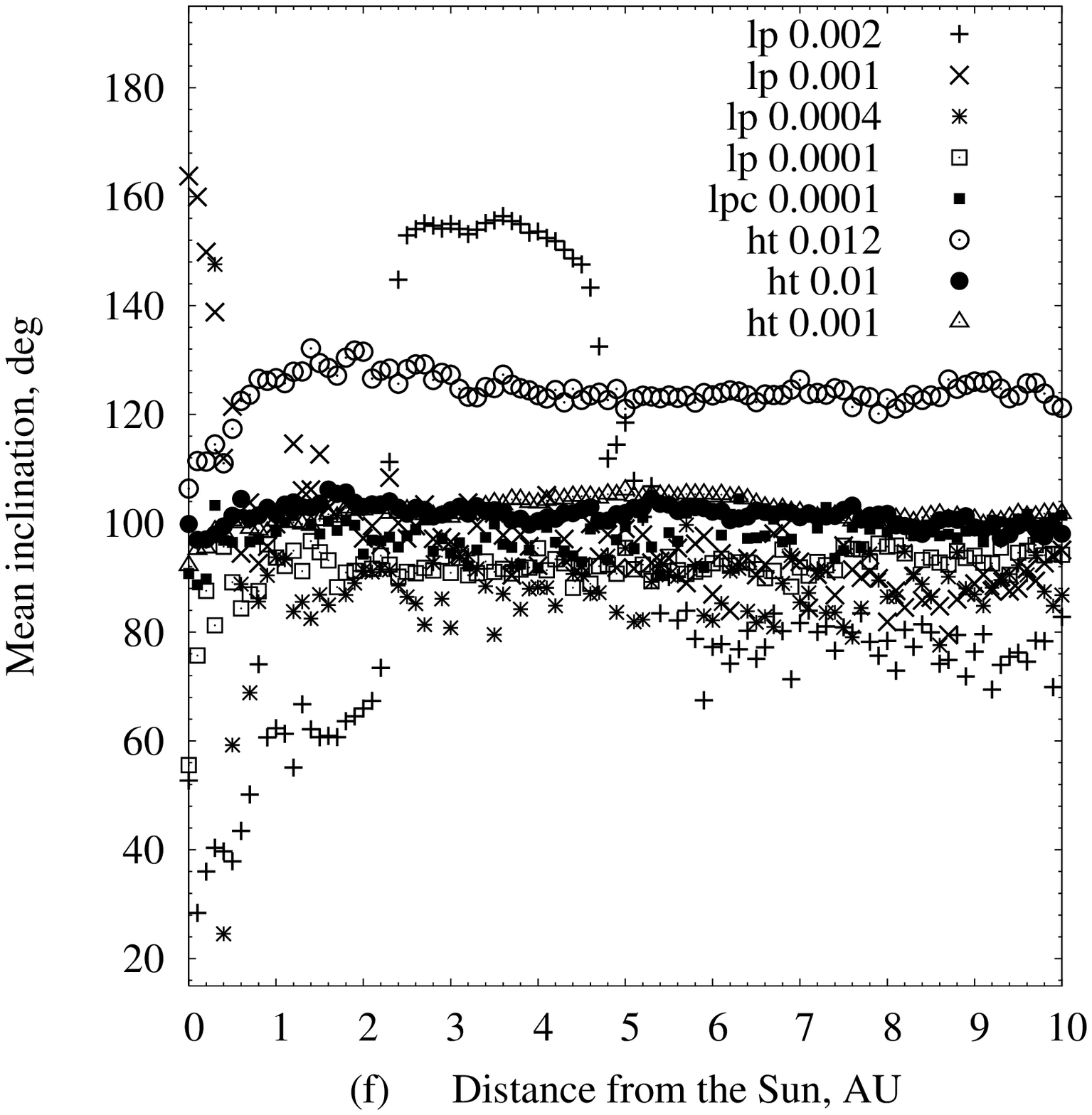} 

\caption{Ipatov et al., Zodiacal cloud...}
\end{figure}%

\begin{figure}    

\includegraphics[width=54mm]{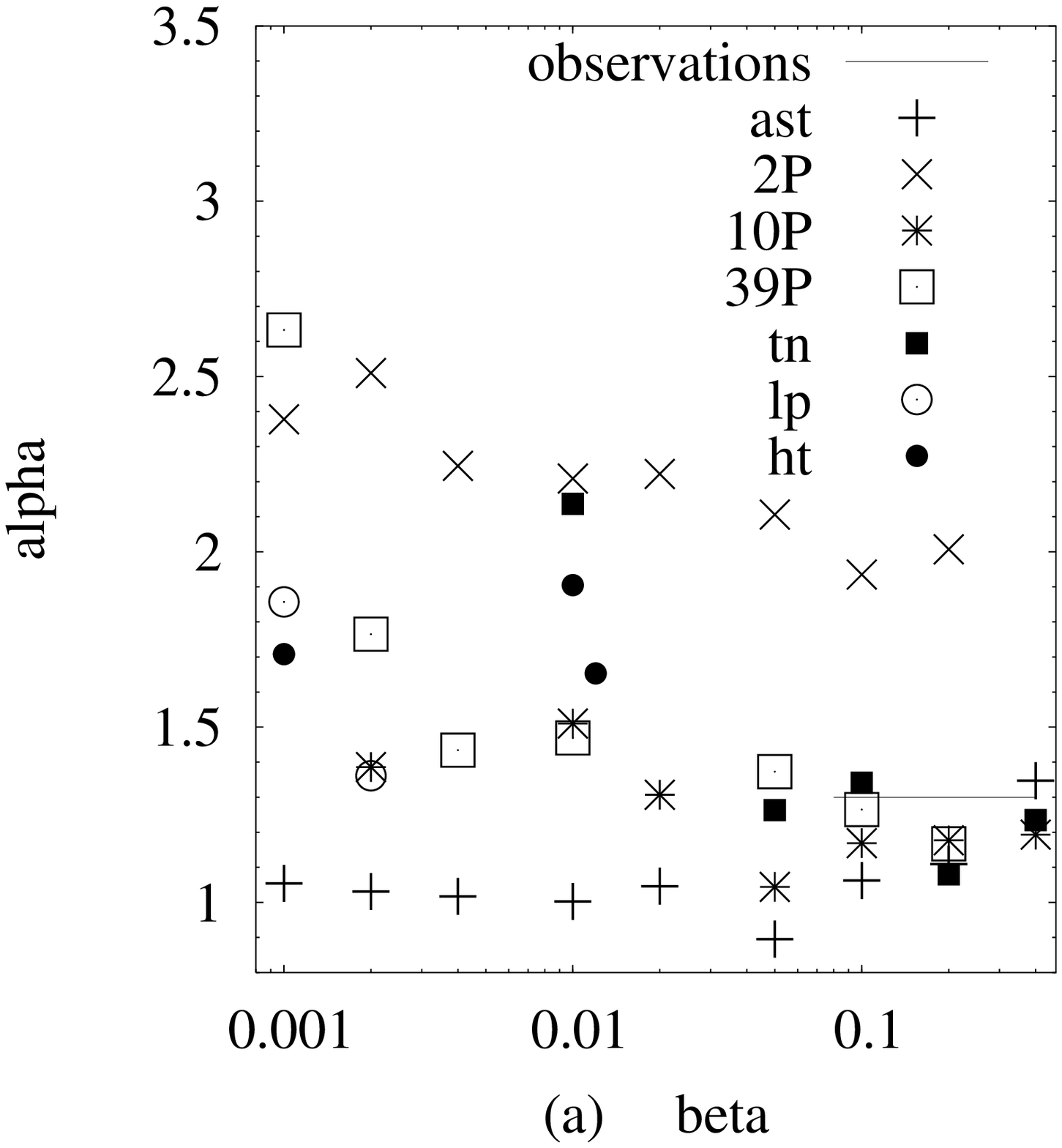}  
\includegraphics[width=54mm]{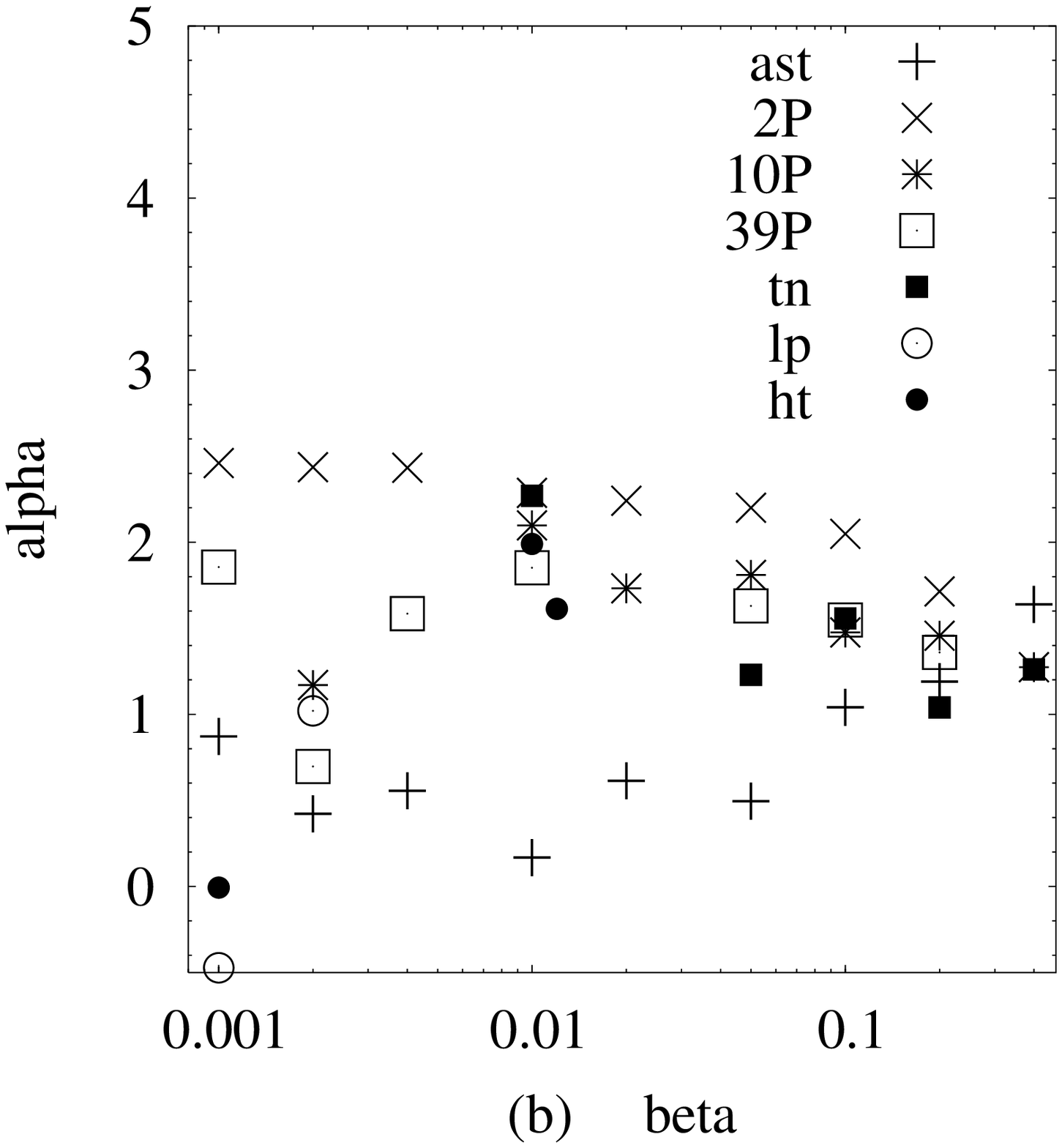}  
\includegraphics[width=54mm]{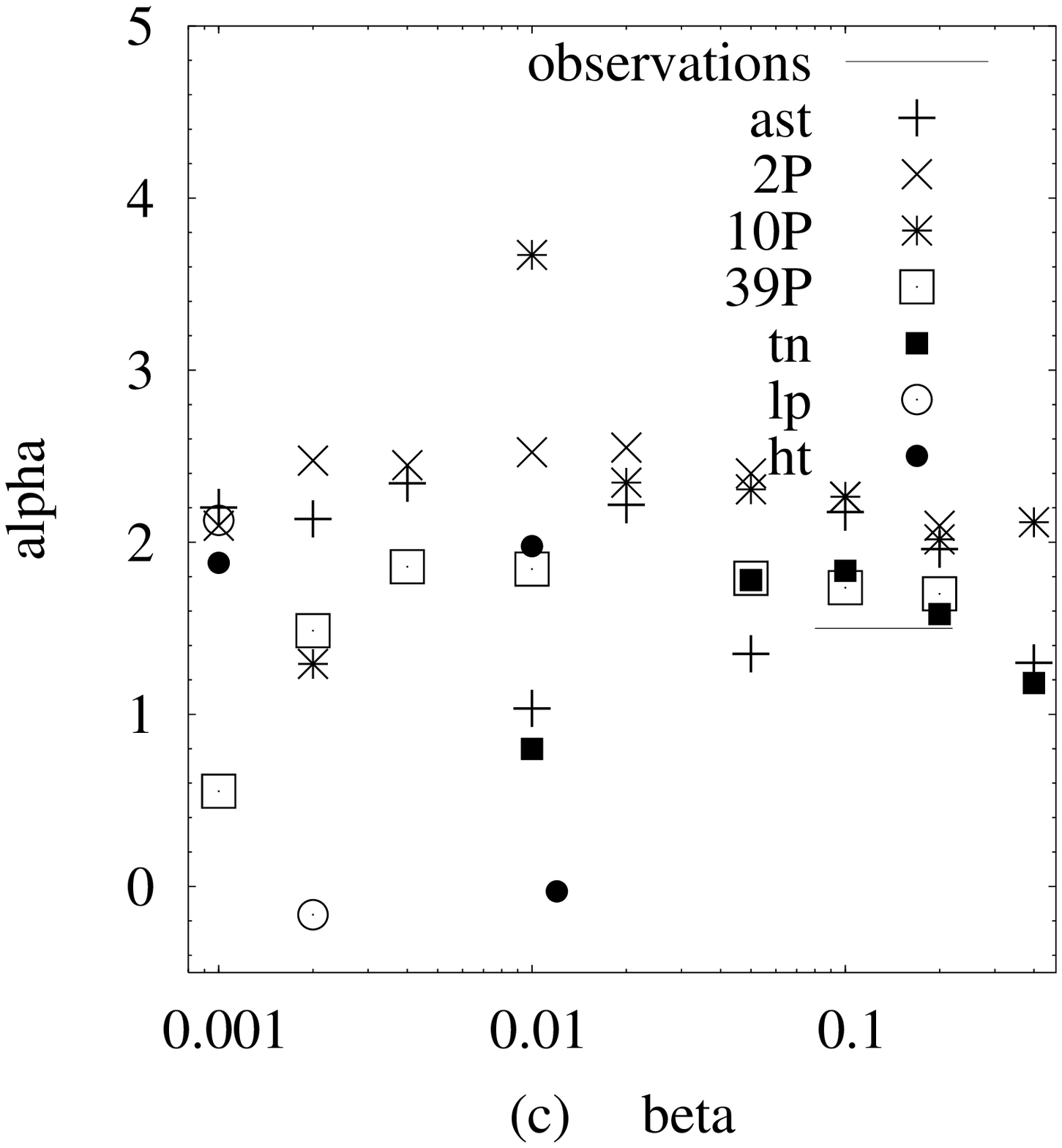}  

\caption{Ipatov et al., Zodiacal cloud...
}
\end{figure}%

\end{document}